\documentclass[lettersize,journal]{IEEEtran}
\usepackage{amsmath,amsfonts}
\usepackage{algorithmic}
\usepackage{algorithm}
\usepackage{array}
\usepackage{subcaption}
\usepackage{textcomp}
\usepackage{stfloats}
\usepackage{url}
\usepackage{verbatim}
\usepackage{graphicx}
\usepackage{cite}
\usepackage{booktabs}
\usepackage{multirow}
\usepackage{url}
\newcommand{\bbdelta}{\boldsymbol{\delta}}
\newcommand{\ccalC}{\mathcal{C}}
\newcommand{\ccalD}{\mathcal{D}}

\begin{document}

\title{Epidemic Forecasting and Behavioral Response Estimation with a Behavior-Coupled SEIR Model}

\author{Sungi Hong, Soham Das, Hossein Khazaei, Alfredo Garcia, Keith Paarporn, and Ceyhun Eksin
\thanks{This work was supported in part by NSF CAREER Award 2239410 and ONR Award N000142612007. The code and data used in simulations are available online at 
\protect\url{https://github.com/SUNGIHONG/Epidemic}.}
\thanks{S. Hong and C. Eksin are with the Department of Industrial and Systems Engineering, Texas A\&M University, College Station, TX 77843 USA (e-mail: eksinc@tamu.edu).}
\thanks{A. Garcia is with the Systems and Information Engineering, University of Virginia, VA 22904.}
\thanks{S. Das is with the Department of Industrial and Systems Engineering, University of Tennessee, Knoxville, TN USA.}
\thanks{H. Khazaei is with Amazon.com, Inc., Seattle, WA USA.}
\thanks{K. Paarporn is with the Department of Computer Science, University of Colorado Colorado Springs, Colorado Springs, CO USA.}
\thanks{This work has been submitted to the IEEE for possible publication.
    Copyright may be transferred without notice, after which this version may no longer be accessible.}
}

\markboth{Preprint submitted to IEEE Transactions on Computational Social Systems}%
{Hong \MakeLowercase{\textit{et al.}}: A Behavior-Coupled SEIR Model for Epidemic Forecasting}

% % column for its text to clear the IEEEpubid mark.

\maketitle

\begin{abstract}
The COVID-19 pandemic has exhibited multiple waves of infections, reflecting feedback between disease transmission and adaptive population behavior. Accurately predicting and understanding these dynamics remains a central challenge for epidemic modeling. In this paper, we develop a forecasting framework based on a Susceptible–Exposed–Infectious–Recovered (SEIR) model coupled with an evolutionary model of behavioral adaptation, in which population compliance with protective measures evolves in response to perceived infection risk and associated costs. We formulate forecasting as the estimation of behavioral incentive parameters that govern the coupled behavior–disease dynamics, using hospitalization time series data together with standard assumptions on disease-related parameters. 
Using the inferred parameters, we generate rolling monthly forecasts of infection levels and evaluate the approach on both synthetic data and hospitalization time series from multiple U.S. states (Texas, California, Colorado, and Arizona). The proposed framework provides a principled way to incorporate behavior–disease feedback into epidemic forecasting while yielding interpretable estimates of population-level incentives. 

\end{abstract}


\section{Introduction}
Recurrent waves and resurgences observed during the COVID-19 pandemic highlight the role of adaptive population behavior in shaping epidemic dynamics~\cite{west2020applying}. An individual's willingness to cooperate with recommended public health measures has a direct impact on disease spread. However, the incentive to cooperate depends on the perceived risks of the disease and the cost of cooperation. As a result, individual incentives evolve dynamically in response to the state of the epidemic. If enough individuals comply, the disease may enter a decline phase before herd immunity is reached; however, the resulting reduction in cases can weaken the incentive to maintain protective behavior, leading to decreased compliance and subsequent resurgence. Consistent with this feedback, the declining effectiveness of public health measures over successive waves has been linked to pandemic fatigue, i.e., waning risk perception and adherence~\cite{du2022pandemic}.

The unpredictable nature of population responses presents a key challenge for epidemic forecasting. Evaluations of the U.S.\ COVID-19 and influenza forecasting hubs found that forecast accuracy varied considerably across models, locations and time, and degraded at longer horizons~\cite{cramer2022evaluation,lopez2024challenges,mathis2024evaluation}. For example, COVID-19 case forecasts did not anticipate the rapid increase at the onset of the winter 2020 wave~\cite{lopez2024challenges}. Machine learning approaches---including deep sequence and graph neural networks, hybrids in which epidemiological dynamics guide neural-network training, and models that fuse mobility and other auxiliary signals---have since aimed to increase the flexibility of data-driven forecasting~\cite{rodriguez2024machine,rodriguez2023einns,ntemi2022autoregressive,yu2022pandora,burnwal2024co,diao2025physics}. Some couple information diffusion with disease spread to represent protective responses~\cite{diao2025physics}. However, these approaches generally do not estimate an explicit, interpretable incentive structure governing behavior. More broadly, a review of 276 COVID-19 intervention-modeling studies published between February 2020 and February 2023 found that only 38\% attempted to incorporate human behavior, often through a single factor such as compliance or mobility~\cite{lee2025review}. Ignoring behavior change can systematically bias forecasts~\cite{eksin2019systematic}, and an analysis of the COVID-19 death projections collected by the U.S.\ Centers for Disease Control and Prevention found that better long-term accuracy was associated with mechanistic representations of transmission and with endogenous projections of behavioral responses~\cite{rahmandad2022enhancing}.

Indeed, a growing body of work explicitly models the interplay between behavior and epidemic spread~\cite{funk2009spread,funk2010modelling,weitz2020awareness,proverbio2026recent}. Game-theoretic and evolutionary formulations, in particular, treat protective behavior as the outcome of individual incentives shaped by infection risk~\cite{huang2022game}. They have been used to analyze coupled behavior--disease dynamics in populations and networks~\cite{paarporn2023sis,frieswijk2022mean,satapathi2022epidemic,saad2023dynamics}, to study strategic mitigation decisions and the design of incentive mechanisms~\cite{certorio2022epidemic,martins2023epidemic,hota2023learning}, and to model vaccination decisions~\cite{dai2025epidemic}, complementing models of opinion- and information-coupled spreading~\cite{she2022networked,an2024coupled}. These studies reveal a range of complex dynamics---from stable endemic states to sustained or growing oscillations---and the conditions under which they arise~\cite{khazaei2021disease,weitz2020awareness}. However, forecast validation remains less common than calibration. In a 2024 scoping review of COVID-19 models with endogenous behavior, 68\% of studies calibrated their models to observed data, but only 25\% compared forecasts with observed data~\cite{hamilton2024incorporating}.

Bringing behavior--disease feedback into data-driven inference raises its own challenges. Behavioral parameters can be difficult to identify from epidemic time series, particularly early in an outbreak~\cite{osi2024parameter}, and different behavioral formulations can fit the same data comparably well while implying different epidemic dynamics, e.g., different reproduction-number trajectories~\cite{gozzi2025comparative}. Recent studies have calibrated behavior-coupled models to COVID-19 data, fitting disease and behavioral data from eight countries with in- and out-of-sample evaluation~\cite{osi2025simultaneous}, calibrating an evolutionary-game model of isolation decisions to local surveillance data~\cite{zai2026behavior}, and inferring time-varying payoff differences for vaccination decisions from epidemiological time series~\cite{maldera2026evolutionary}. Complementing these efforts, we focus on a latent protective behavior, social distancing, whose frequency-dependent payoff structure---summarized by four payoff gaps---is estimated from hospitalization data, linked to the qualitative dynamical regimes of the coupled system, and used for out-of-sample forecasting.

In this paper, we propose an epidemic forecasting framework based on a mechanistic behavior-coupled SEIR model studied in prior work~\cite{khazaei2021disease}. While prior work focused on the qualitative dynamics of such models, we leverage the framework for data-driven inference and prediction. The model combines a standard SEIR structure governing disease transmission with replicator dynamics that capture how individuals adjust their behavior in response to infection risk and the incentives for compliance~\cite{EksinWeitz16}. That is, population behavior modulates the transmission rate, while infection prevalence shapes the payoffs of complying with protective measures. Treating behavior as a latent, dynamically evolving quantity governed by incentives, we estimate the payoff parameters from infection trajectories, assuming that disease parameters are available from the literature. We demonstrate the framework on data from four U.S.\ states (Texas, California, Colorado, and Arizona) that exhibit diverse epidemic trajectories.

The main contributions of this paper are threefold: (i) a rolling forecasting framework that infers time-varying population-level behavioral incentives from infection time series via weighted nonlinear least squares; (ii) an interpretation linking the current inferred incentives to the qualitative dynamical regimes of the coupled system, relating them to distinct behavioral--epidemic patterns; and (iii) an evaluation of the forecasting framework on synthetic data and four U.S.\ states showing improved short- to medium-term forecast accuracy over a standard seasonal SEIR baseline.

The remainder of this paper is organized as follows.
Section~\ref{sec:model} introduces the behavior-coupled SEIR model and the payoff-estimation and forecasting framework. Section~\ref{sec:synthetic_section} describes the synthetic experiments and the results.  Section~\ref{sec:hospitalization_section} describes the estimation process using hospitalization data, and reports the forecasting results for multiple states including a comparison with a standard seasonal SEIR baseline. Section~\ref{sec:conclusion} discusses limitations and potential extensions.

\section{Epidemic Forecasting Framework} \label{sec:model}
\subsection{Behavior-Coupled SEIR Model} 

The population consists of individuals in susceptible ($s$), exposed ($e$), infected ($i$), and recovered ($r$) states (compartments) where $s$, $e$, $i$, and $r$ denote the fraction of individuals in each state. Upon contracting the disease, a susceptible individual moves to the exposed state. After the incubation period $(\frac{1}{\eta})$, the exposed individual becomes infected. Infected individuals can transmit the disease to susceptible individuals. The infectious individuals eventually reach the recovered state after the infection period ($\frac{1}{\alpha}$). Assuming a well-mixed population, the fraction of susceptible individuals that move to the exposed state is determined by an infection rate $\beta(t)$ and the contact rate captured by $s\cdot i$. The SEIR dynamics are as follows,
\begin{align}
    \dot s &= -\beta(t) \cdot s\cdot i \label{eq_s}\\
    \dot e &= \beta(t) \cdot s \cdot i - \eta \cdot e \label{eq_e}\\
\dot i &= \eta \cdot e - \alpha \cdot i \label{eq_i}\\
\dot r &= \alpha \cdot i \label{eq_r}.
\end{align}
In the standard SEIR model, the infection rate $\beta(t)$ is either constant or varies according to seasonality \cite{tamerius2013environmental}. Here, the infection rate depends on the population behavior $x\in[0,1]$ representing the fraction of the society cooperating with social distancing guidelines, 
\begin{equation} \label{eq_beta}
    \beta(t) = \beta_0(t) (1-x)
\end{equation}
where $\beta_0(t)$ is the infection rate of the disease subject to seasonality. Setting $x\equiv 0$ (no behavioral coupling) reduces \eqref{eq_beta} to $\beta(t)=\beta_0(t)$, recovering the standard seasonal SEIR model \eqref{eq_s}--\eqref{eq_r}, which we use as a baseline in Section~\ref{sec:seir_comparison}.

The incentive to comply with the recommended social distancing measures, e.g., wearing a mask, is higher if the disease is more prevalent, i.e., $i$ is large, in the population. Conversely, the incentive to comply is lower if the disease is less prevalent, i.e., $i$ is small, in the population. We reflect this relation between infection risks and incentives using an environment parameter $\theta:=1-\epsilon i$ for $\epsilon>0$, and the following mapping from environment to the payoff matrix,
\begin{align}\label{eq_payoffs}
    A_\theta = \theta  \underbrace{\begin{bmatrix} R_1 & S_1 \\ T_1 & P_1 \end{bmatrix}}_{\text{$A_1$}}  + (1-\theta) \underbrace{\begin{bmatrix} R_0 & S_0 \\ T_0 & P_0 \end{bmatrix}}_{\text{$A_0$}} ,
\end{align}
which has been used in feedback-evolving models ~\cite{EksinWeitz16,tilman2020evolutionary,khazaei2021disease,paarporn2023sis}.
The constant $\epsilon$ translates the current disease prevalence to a state of the environment, and can be interpreted as the sensitivity of the population to the current disease prevalence. The payoff matrix $A_\theta$ \eqref{eq_payoffs} captures the payoffs of two players with two actions ($\mathcal{C}$--{\it Cooperate}, $\mathcal{D}$--{\it Defect}) in a symmetric game. 
The elements in the first row of $A_\theta$ ($R_\theta$ or $S_\theta$) are the payoffs that a cooperating row player attains against another cooperator or defector in the population. The elements in the second row of $A_\theta$ ($T_\theta$ or $P_\theta$) are the payoffs that a defector attains against another cooperator or defector in the population.

In a good environment $\theta=1$ ($i=0$), we assume the payoff matrix $A_1$ corresponds to a Prisoner's Dilemma game such that $\delta_{RT_1}:=R_1-T_1<0$ and $\delta_{SP_1}:=S_1-P_1<0$. In a Prisoner's Dilemma game, the unique rational behavior (Nash equilibrium) is defection by both players. This represents the situation where the individuals do not need to comply with social distancing measures (cooperate) if the disease is not present in the population. 

The fraction of cooperators $x$ in the population increases if a cooperator's payoff $u_\ccalC(x,\theta)$ is larger than that of defectors $u_\ccalD(x,\theta)$. Specifically, the behavior follows replicator dynamics\footnote{The manner in which agents interact and revise their strategies is called a revision protocol \cite{sandholm2010population}. Many distinct types of revision protocols, e.g., pairwise comparison, induce the replicator dynamics. For agents to perform these revision protocols, it is not necessary for them to physically interact with one another in order to learn about others' payoffs -- these may be learned through information sources such as media and social networks. Thus, we assume the interactions that underlie the replicator equation do not factor into the physical epidemic spreading process.}, 
\begin{equation} \label{eq_replicator}
    \dot x = x (1-x) (u_\ccalC(x,\theta) - u_\ccalD(x,\theta)),
\end{equation}
where 
\begin{align}
    u_\ccalC(x,\theta) &= x R_\theta  + (1-x) S_\theta \label{eq_cooperate_payoff}\\
    u_\ccalD(x,\theta) &= x T_\theta + (1-x) P_\theta \label{eq_defect_payoff}
\end{align}
where $R_\theta$, $S_\theta$, $T_\theta$, and $P_\theta$ are the corresponding entries of $A_\theta$ in \eqref{eq_payoffs}. The payoffs above respectively represent the payoffs of a cooperator ($\ccalC$) and a defector ($\ccalD$) randomly meeting another player from the population with $x$ cooperators.

Depending on the gap between payoff parameters in $A_0$, i.e., $\delta_{RT_0}:=R_0-T_0$ and $\delta_{SP_0}:=S_0-P_0$, the infection prevalence over time can display different transient dynamics, as demonstrated in prior work~\cite{khazaei2021disease}.
Whenever mutual defection from compliant behavior is rational in $A_0$ ($\delta_{SP_0} \leq 0$), the dynamics yield a single concentrated peak of infections, followed by exponential decay to zero. This is intuitive, since the incentives in both states $A_0$ and $A_1$ do not sustain compliant behavior. We label these as ``tragedy of the commons (TOC)'' regimes~\cite{hardin2009tragedy}. The specific labels are Q2-TOC and Q3-TOC, depending on whether $\delta_{RT_0}>0$ or $\delta_{RT_0}<0$, respectively.

When mutual defection is not the rational behavior in $A_0$ ($\delta_{SP_0} > 0$), the dynamics yield infectious oscillations whose peaks are less severe than the peak in the TOC regimes. When $A_0$ is an anti-coordination game ($\delta_{SP_0} > 0$, $\delta_{RT_0}  \leq 0$), the oscillations of $i_t$ diminish in amplitude. Hence, we name this region Q4-Diminishing Oscillations. When cooperation is the dominant action of $A_0$ ($\delta_{SP_0} > 0$, $\delta_{RT_0} > 0$), the oscillations are either diminishing or initially growing in amplitude, depending on the value of $\delta_{RT_0}$. We label these regions Q1-Diminishing Oscillations and Q1-Oscillating TOC, respectively---see Appendix~\ref{Append: A} and Figure \ref{fig:AllDynamics} for the specific regions.

% 
% A main challenge in developing a prediction models is to find a mechanism that reflect the back-and-forward dynamics between society and the disease. Such dynamics display themselves as different waves of the disease. 
% This paper focuses on the dynamics between the disease and the behavior of the paper. 
\subsection{Weighted Least-Squares for Estimating Payoff Matrices}
The payoff matrices that govern behavioral dynamics are generally unknown and should be inferred from epidemic data. We consider a parameter estimation framework based on weighted least-squares to learn the payoff parameters from observed infection trajectories.

Define the unknown payoff gaps
$
{\bbdelta}
=
\left[
\delta_{RT_1},\,
\delta_{RT_0},\,
\delta_{SP_1},\,
\delta_{SP_0}
\right]^\top
\in \mathbb{R}^4$. 
%where $\delta_{RT_1} = R_1-T_1$, $\delta_{RT_0} = R_0 - T_0$, $\delta_{SP_1} = S_1 - P_1$, and $\delta_{SP_0} = S_0 - P_0$
 Rather than estimating the payoff parameters in \eqref{eq_payoffs}, it suffices to estimate $\boldsymbol{\delta}$. To see this, note that we can express the replicator dynamics in \eqref{eq_replicator} using $\boldsymbol{\delta}$,
\begin{align}
     \dot x = x (1-x)& \cdot \bigg[x \theta \delta_{RT_1} + x (1-\theta) \delta_{RT_0}  \nonumber\\
    &+ (1-x) \theta \delta_{SP_1} + (1-x) (1-\theta) \delta_{SP_0} \bigg]. \label{eq_rep_expanded}
\end{align}
We get \eqref{eq_rep_expanded} by substituting for $R_\theta = \theta R_1 + (1-\theta) R_0$, $S_\theta = \theta S_1 + (1-\theta) S_0$, $T_\theta = \theta T_1 + (1-\theta) T_0$, and $P_\theta = \theta P_1 + (1-\theta) P_0$ in $u_\ccalC$ \eqref{eq_cooperate_payoff} and $u_\ccalD$  \eqref{eq_defect_payoff} and then expanding the $u_\ccalC - u_\ccalD$ term in \eqref{eq_replicator}.

% Based on $\dot x$, the coupled seir–replicator dynamics are given by
% \begin{align}
% \dot s &= -\beta_0 (1-x) s i + \mu r, \\
% \dot e &= \beta_0 (1-x) s i - \eta e, \\
% \dot i &= \eta e - \alpha i, \\
% \dot r &= \alpha i - \mu r \label{eq:rdot}
% \end{align}
Given a candidate parameter vector $\boldsymbol{\delta}$, we numerically solve the above ODE system using an adaptive Runge--Kutta method to obtain the model-generated infection trajectory $i_{(t;\boldsymbol{\delta})}$, where $i_{\text{obs}(t)}$ denotes the observed infection trajectory. The solver uses internal adaptive time steps for integration and these steps do not correspond to the data timescale. To match the daily observations, we evaluate the resulting trajectory at integer time points $t=0,1, \dots ,T$, which are days. We estimate $\boldsymbol{\delta}$ by solving the weighted nonlinear least-squares problem 
\begin{equation}
\boldsymbol{\widehat \delta}
=
\arg\min_{\boldsymbol{\delta}}
\sum_{t=0}^{T-1}
\gamma^{\,2(t-(T-1))}
\left(
i_{(t;\boldsymbol{\delta})}
-i_{\text{obs}(t)}
\right)^2,
\label{eq:nonlinear_wls}
\end{equation}
where $\gamma>1$ emphasizes more recent observations.
To enforce $\delta_{RT_1}<0$ and $\delta_{SP_1}<0$, we reparametrize
$\delta_{RT_1}=-\operatorname{softplus}(z_{RT_1})$ and
$\delta_{SP_1}=-\operatorname{softplus}(z_{SP_1})$,
where $\operatorname{softplus}(z)=\log(1+e^z)$.

% \section{Experimental Results} \label{sec:results}

\section{Forecasting using synthetic data} \label{sec:synthetic_section}
\subsection{Setup} 
\label{sec:synthetic_setup}

We generate synthetic infection data by simulating the model dynamics \eqref{eq_s}-\eqref{eq_r}, and \eqref{eq_rep_expanded} for $T$ steps and then corrupting it with additive Gaussian noise,  \(
i_{\text{syn}(t)} = i_{(t)} + \mathcal{N}(0,\, \sigma^2),\) with $\sigma^2=10^{-4}$ for all $t\in[0,T]$.  To evaluate the robustness of the proposed framework to uncertainty in the disease parameters, we perturb the nominal values of $(\alpha,\eta,\beta_0)$ using Gaussian noise with standard deviations equal to $10\%$ of their respective nominal values, clipped within
$\pm20\%$ of the nominal values. The corresponding parameter ranges are summarized in Table~\ref{varpar}.

\begin{table}[h] 
\centering
\caption{{Parameter values and uncertainty ranges used in the synthetic experiments}} \label{varpar}
\begin{tabular}{c|c}
Parameter (Nominal value) & Uncertainty Range \\ \hline
$\beta_0 = 0.7$     & $[0.56,\;0.84]$ \\
$\alpha = 0.15$    & $[0.12,\;0.18]$ \\
$\eta = 0.1428$    & $[0.114,\;0.171]$ \\
$\epsilon  = 20$    & 20 (Fixed) \\
$\gamma  = 1.01$    & 1.01 (Fixed) \\\hline
\end{tabular}
\end{table}

We assume the data collection starts at $t=0$ and goes for $T=60$ days.  We consider six scenarios corresponding to all possible dynamic patterns arising from different payoff values in $A_0$, namely Q1-Diminishing Oscillations, Q1-Oscillating TOC, Q1-Boundary, Q2-TOC, Q3-TOC, and Q4-Diminishing Oscillations---see Appendix ~\ref{Append: A} for definitions. The ground-truth payoff gaps used to generate each scenario are listed in Table~\ref{tab:synthetic_delta}. For each behavior regime (scenario), we conduct $N=200$ simulations and present the confidence intervals indicating the $2.5$th and $97.5$th percentiles of the simulated trajectories. \begin{table}[t]
\centering
\small
\caption{Ground-truth payoff gaps
$\bbdelta=[\delta_{RT_1},\delta_{RT_0},\delta_{SP_1},\delta_{SP_0}]$
used to generate the synthetic scenarios.}
\label{tab:synthetic_delta}
\begin{tabular}{lcccc}
\hline
Scenario & $\delta_{RT_1}$ & $\delta_{RT_0}$ & $\delta_{SP_1}$ & $\delta_{SP_0}$ \\ \hline
Q2-TOC                       & $-0.50$ & $1.40$  & $-0.25$ & $-2.40$ \\
Q3-TOC                       & $-0.50$ & $-2.00$ & $-0.25$ & $-1.40$ \\
Q4-Diminishing Oscillations  & $-0.50$ & $-0.20$ & $-0.25$ & $0.75$  \\
Q1-Diminishing Oscillations  & $-0.50$ & $0.20$  & $-0.25$ & $1.00$  \\
Q1-Oscillating TOC           & $-0.50$ & $1.50$  & $-0.25$ & $0.50$  \\
Q1-Boundary                  & $-0.50$ & $1.20$  & $-0.25$ & $0.60$  \\ \hline
\end{tabular}
\end{table}

\begin{figure*}[h]
\centering

% --- Row 1 ---
\begin{subfigure}[t]{0.31\textwidth}
  \centering
  \includegraphics[width=\linewidth]{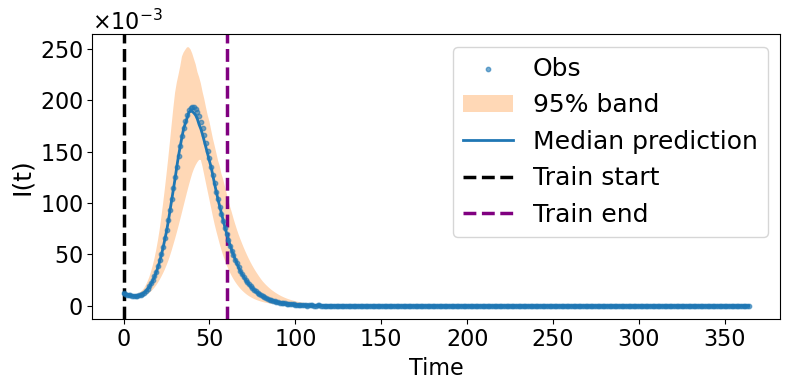}
  \caption{Q2-TOC (95\% CI)}
\end{subfigure}\hfill
\begin{subfigure}[t]{0.31\textwidth}
  \centering
  \includegraphics[width=\linewidth]{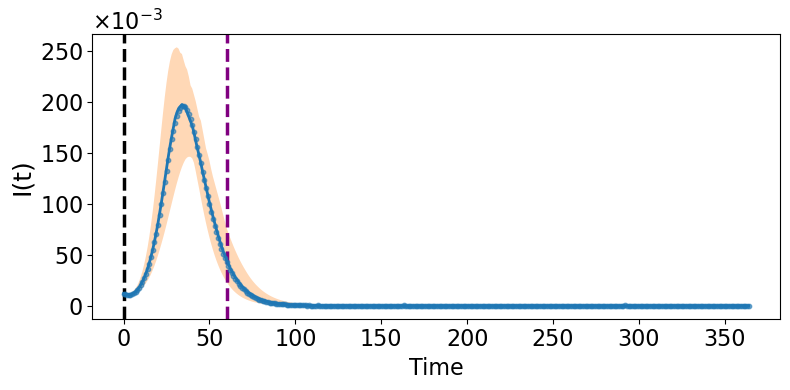}
  \caption{Q3-TOC (95\% CI)}
\end{subfigure}\hfill
\begin{subfigure}[t]{0.31\textwidth}
  \centering
  \includegraphics[width=\linewidth]{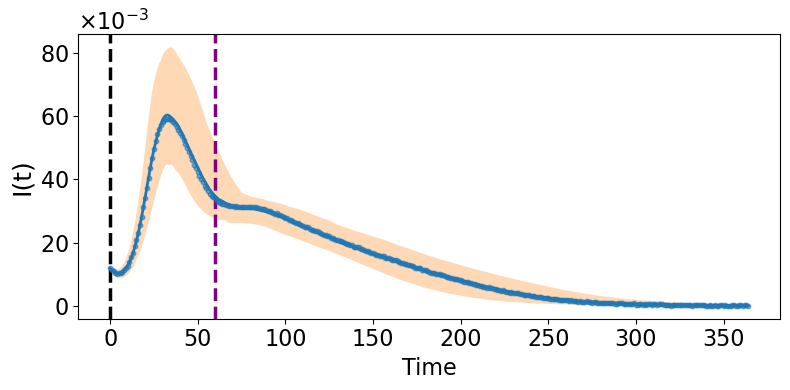}
  \caption{Q4-Diminishing Oscillations (95\% CI)}
\end{subfigure}

\vspace{2mm}

% --- Row 2 ---
\begin{subfigure}[t]{0.31\textwidth}
  \centering
  \includegraphics[width=\linewidth]{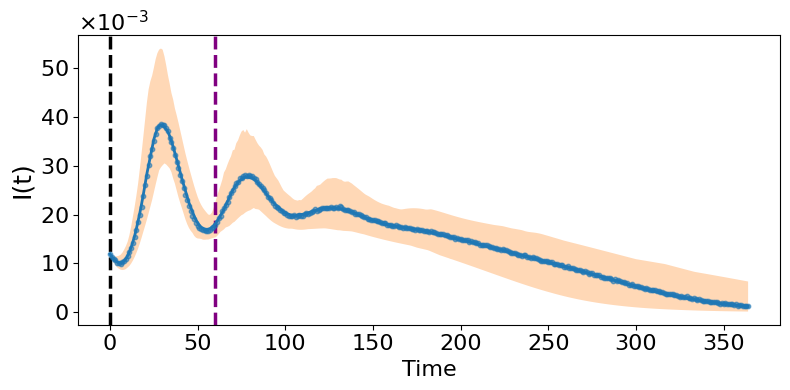}
  \caption{Q1-Diminishing Oscillations (95\% CI)}
\end{subfigure}\hfill
\begin{subfigure}[t]{0.31\textwidth}
  \centering
  \includegraphics[width=\linewidth]{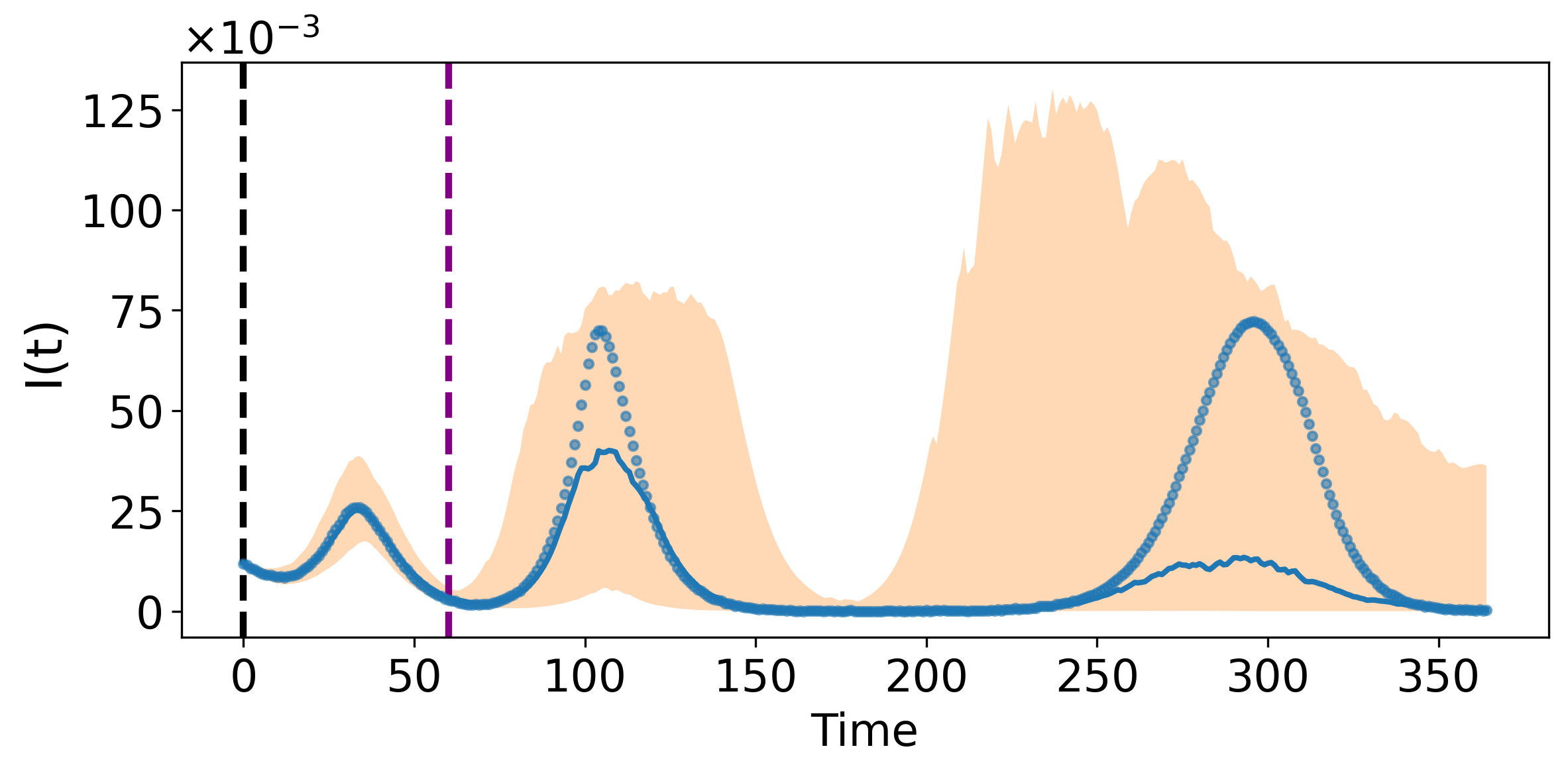}
  \caption{Q1-Oscillating TOC (95\% CI)}
\end{subfigure}\hfill
\begin{subfigure}[t]{0.31\textwidth}
  \centering
  \includegraphics[width=\linewidth]{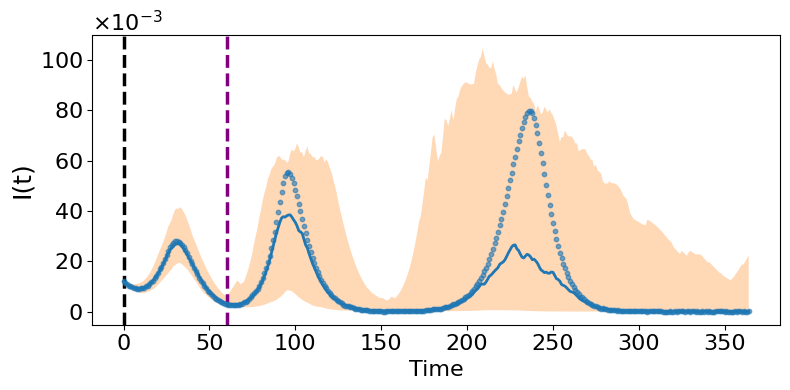}
  \caption{Q1-Boundary (95\% CI)}
\end{subfigure}

\caption{Benchmark prediction results across regimes. All panels show predictions with 95\% confidence intervals under $\pm 20\%$ perturbations of disease-related parameters.}
\label{fig:benchmark_6panel}

\end{figure*}

\subsection{Synthetic benchmark results} \label{syn_res}
To illustrate the estimation performance, we report the forecasting results for a representative training horizon $T=60$ in Figure \ref{fig:benchmark_6panel}, and summarize the corresponding estimation accuracy in Table~\ref{tab:synthetic_summary}. For each scenario, Table~\ref{tab:synthetic_summary} reports  the median residual norm $\operatorname{median} \!\left(\|r^{(k)}\|_2\right)$ and the median $L^1$ error in the payoff-gap estimates, $\operatorname{median}(\|\widehat{\bbdelta}^{(k)} -\bbdelta^{\mathrm{true}}\|_1)$, where $\widehat{\bbdelta}^{(k)}$ is the estimate obtained from the $k$-th Monte Carlo sample.

For the behavior regimes Q2-TOC, Q3-TOC, and Q4-Diminishing Oscillations, the confidence intervals are narrow and closely track the infection trajectories. For the Q1-Oscillating TOC regime and the Q1-Boundary, the median trajectory can miss the observed peaks.  We observe that the payoff parameter estimates are still close to the ground truth in these regimes (Table~\ref{tab:synthetic_summary}); small estimation errors can lead to significant deviations in the resulting dynamics when the actual payoffs are near the boundaries of different regimes. Nevertheless, across these synthetic benchmarks, the confidence bands consistently cover the true infection trajectories.

\begin{table}[t]
\centering
\caption{Estimation accuracy for the synthetic benchmarks
($T_{\text{start}}=0$, $T_{\text{train}}=60$).}
\label{tab:synthetic_summary}
\resizebox{\columnwidth}{!}{%
\begin{tabular}{lcc}
\hline
Scenario
& $\operatorname{median}\!\left(\|r^{(k)}\|_2\right)$
& $\operatorname{median}\!\left(\left\|\widehat{\bbdelta}^{(k)}-\bbdelta^{\mathrm{true}}\right\|_1\right)$ \\ \hline
Q3-TOC                       & $5.87\times10^{-4}$ & 3.6419 \\
Q2-TOC                       & $5.80\times10^{-4}$ & 3.0364 \\
Q4-Diminishing Oscillations  & $5.58\times10^{-4}$ & 0.0113 \\
Q1-Diminishing Oscillations  & $5.71\times10^{-4}$ & 0.0263 \\
Q1-Oscillating TOC           & $5.67\times10^{-4}$ & 0.1109 \\
Q1-Boundary                  & $5.73\times10^{-4}$ & 0.0950 \\ \hline
\end{tabular}}
\end{table}

\section{Forecasts Based on Hospitalization Data} \label{sec:hospitalization_section}

\subsection{Forecasting Setup: Hospitalization Data During COVID-19 and Disease Parameters}
We validate the estimation method leveraging the SEIR model coupled with population dynamics using hospitalization data within the first 12 months of the COVID-19 pandemic from Texas, California, Colorado, and Arizona~\cite{CovidData}.
Hospitalizations due to COVID-19 in a given locality is one of the most reliable measurements available during the pandemic, while the number of daily infected was subject to various ascertainment biases and availability of testing \cite{buitrago2022studies}.
We estimate the daily number of infected from hospitalization data, assuming that a fraction $\kappa$ of infected individuals is hospitalized $\tau$ days later on average,
\begin{equation}\label{eq_hospitalized}
    h_t = \kappa \, i_{t-\tau}.
\end{equation}
Given hospitalization counts $h_{0:T}$ and estimates of $\kappa$ and $\tau$, we recover $\hat{i}_t = h_{t+\tau}/\kappa$ after smoothing the series with a Savitzky--Golay filter; the full procedure is given in Algorithm~\ref{alg:two}. Here, the population size is used to compute the infection proportion $i_t$---see Appendix~\ref{parasel} for specific values.

\begin{algorithm}
\caption{Estimating the number of infected using hospitalization data}
\label{alg:two}

\begin{algorithmic}

\STATE \textbf{Input:}
$h_{\tau:T+\tau}$, $\kappa$, $\tau$

\STATE \textbf{Output:}
$\{\hat{i}_t\}_{t=0}^{T}$ and
$\{\hat{\dot{i}}_t\}_{t=1}^{T}$

\STATE Smooth hospitalization data using a Savitzky--Golay filter;
see \cite{schafer2011savitzky}

\FOR{$t=0,\ldots,T$}
    \STATE $\hat{i}_t \gets \dfrac{h_{t+\tau}}{\kappa}$
\ENDFOR

\FOR{$t=1,\ldots,T$}
    \STATE $\hat{\dot{i}}_t
    \gets \hat{i}_t-\hat{i}_{t-1}$
\ENDFOR

\RETURN $\{\hat{i}_t\}_{t=0}^{T},
\{\hat{\dot{i}}_t\}_{t=1}^{T}$

\end{algorithmic}
\end{algorithm}

Disease-related parameters are generally less sensitive to external factors, as they are  mainly intrinsic to the disease or depend on the demographic characteristics of the population. Unless there is a significant external factor such as vaccination or mutation, they remain relatively stable over time. Accordingly, we use ranges of values of the disease-related parameters established in the literature for COVID-19 in the first 12 months. 
Specifically, we assume that the inherent disease transmission rate of COVID-19 is subject to seasonal changes. In particular, we have 
\begin{equation}
   \beta_0(t) = \frac{\beta_0^s + \beta_0^w}{2} + \frac{\beta_0^w-\beta_0^s}{2}\cos{\frac{2\pi(t-t_0)}{365}}
\end{equation}
where $\beta_0^w$ and $\beta_0^s$ denote the baseline transmission rates during winter and summer, respectively, as commonly used in the literature \cite{dushoff2004dynamical, tamerius2013environmental, greenman2004external,altizer2006seasonality}. The parameter $t_0$ represents a phase shift that determines the timing of the seasonal peak in transmission, i.e., the time at which $\beta_0(t)$ reaches its maximum value $\beta_0^w$.

The disease-related parameters considered in our model include 
$\beta_0^s$, $\beta_0^w$, $\eta$, $\alpha$, $\kappa$, and $\tau$.  For each state (e.g., Texas, California, Colorado, and Arizona), we adopt nominal disease-related parameter values established in the literature. The nominal values and their associated uncertainty ranges are summarized in Table~\ref{tab:SEIRparameterValues}---see Appendix~\ref{parasel} for the literature basis.

\begin{table}[h]
\caption{Summary of parameters and uncertainty ranges}
\label{tab:SEIRparameterValues}
\begin{center}
\begin{tabular}{c c c}
\hline
Parameter & Nominal Value & Uncertainty Range ($\pm 20\%$) \\
\hline
$\eta$ & 0.143 & $[0.1144,\; 0.1716]$ \\
$\alpha$ & 0.15 & $[0.12,\; 0.18]$ \\
$\kappa$ & 0.07 & $[0.056,\; 0.084]$ \\
$\tau$  & $9$ & $\{7,8,9,10\}$ \\
$\epsilon$  & $20$ & $20\ (\text{fixed})$ \\
$\gamma$  & $1.01$ & $1.01\ (\text{fixed})$ \\
\hline
$\beta_0^s$ (AZ, TX) & 0.25 & $[0.20,\; 0.30]$ \\
$\beta_0^w$ (AZ, TX) & 0.55 & $[0.44,\; 0.66]$ \\
\hline
$\beta_0^s$ (CO) & 0.25 & $[0.20,\; 0.30]$ \\
$\beta_0^w$ (CO) & 0.70 & $[0.56,\; 0.84]$ \\
\hline
$\beta_0^s$ (CA) & 0.35 & $[0.28,\; 0.42]$ \\
$\beta_0^w$ (CA) & 0.55 & $[0.44,\; 0.66]$ \\
\hline
\end{tabular}
\end{center}
\footnotesize{\textit{Note:} These parameter choices reflect climatic differences across states. Colder inland climates tend to exhibit stronger seasonal forcing in respiratory virus transmission, whereas warmer coastal regions show weaker seasonal variation \cite{audi2020seasonality,merow2020seasonality}. Accordingly, Colorado exhibits a stronger winter effect due to its colder continental climate, while California shows a relatively smaller seasonal gap compared to the other states. To capture heterogeneity in the timing of seasonal transmission, we set the winter peak timing as follows: Arizona (AZ) and Texas (TX) peak on February 20, Colorado (CO) peaks on December 20, and California (CA) peaks on January 10. These differences reflect regional climate patterns across the United States. In particular, southern states such as Arizona and Texas tend to experience milder winters, with the coldest period occurring later in the season, whereas northern states such as Colorado typically reach their coldest temperatures earlier in winter. California, with its coastal climate, exhibits an intermediate timing of the winter minimum.}
\end{table}

In addition to the disease parameters, we set the behavioral coupling parameter $\epsilon$ to $20$. This choice implies that the environment $\theta=1-\epsilon i$ reaches zero at 5\% infection prevalence. We also tested values of $\epsilon$ in the range $10-40$ and found that inferred behavioral regimes and forecasting results are not sensitive to this choice. 

We make predictions using the estimated payoff parameters $\hat\bbdelta$ by starting the forecasts from the end of the training horizon, with training windows of 90, 120, and 150 days. We construct confidence intervals for our forecasts using Monte Carlo simulations. Specifically, we sample disease-parameter values from their uncertainty ranges, simulate the resulting dynamics forward, and use the distribution of projected trajectories to quantify the forecast uncertainty.

\subsection{Forecast results for Texas, California, Colorado, and Arizona}\label{real} 

In Figures~\ref{fig:TXCA} and~\ref{fig:COAZ}, we present the observed infection trajectories, the median forecast obtained from Monte Carlo simulations, and the corresponding confidence interval of the forecasts for TX, CA, CO, and AZ. We provide additional context by overlaying the plots with the timing of the mitigation policies implemented by each state. We express these policies in three intensity levels: Level 1 (relaxed mitigation), Level 2 (mitigation), and Level 3 (strong suppression)---see Appendix~\ref{sec_mandates} for details.
Fig.~\ref{fig:regime_tx_ca_az_co} tracks the median estimated payoff gaps $\delta_{RT_0}$ and $\delta_{SP_0}$ for all states across training windows of 90, 120, and 150 days.

Across all four states, the proposed forecasting approach closely tracks the infection trajectory over a horizon of at least 60--90 days. As the length of the training window increases from 90 to 150 days, the forecasts become less sensitive to parameter uncertainty, as reflected in narrower confidence bands. 
The estimated payoff gaps and the corresponding behavior regime change with the training window. In particular, California shifts from Q2-TOC to the cooperation-dominant Q1-Diminishing Oscillations regime as the training window increases. Arizona remains in Q1 across all training windows, but transitions from Q1-Oscillating TOC at 90 days to Q1-Diminishing Oscillations at 120 and 150 days. Colorado maintains a stable Q4-Diminishing Oscillations regime with decreasing payoff-gap magnitude, whereas Texas exhibits regime switching across training windows.

\subsubsection{Texas}
Forecasts in TX based on 90, 120, and 150 days of training data capture the ensuing disease trajectory observed---see Fig.~\ref{fig:TXCA}-(a,c,e). In particular, the forecast confidence intervals capture the observed path of the disease. The median forecasts based on 90 days of training data consistently underestimate the infection levels in the state, while the median forecasts based on 120 and 150 days of training data are able to capture both the timing and magnitude of the second peak, as well as the trend and the magnitude of the infection levels observed in the next 4 to 6 months.

Under the 90-day training window, Texas is classified as Q1-Oscillating TOC, with $\delta_{RT_0} > 0$ and $\delta_{SP_0} > 0$, indicating a cooperation-dominant game (Fig.~\ref{fig:TXCA}(a)). Despite the dominance of cooperation in $A_0$, the dynamics belong to the  oscillating TOC regime, which implies a resurgence of the disease. 
The 120-day training period places greater weight on the sustained Level 2 mitigation phase, while still including the earlier relaxed (Level 1) policy phase. This shift in the data window is accompanied by a qualitative change in the inferred game: the system moves from a cooperation-dominant but oscillatory regime to an anti-coordination regime with diminishing oscillations (Q4-Diminishing Oscillations), characterized by $\delta_{RT_0} < 0$ and $\delta_{SP_0} > 0$---see Fig.~\ref{fig:regime_tx_ca_az_co}.  In this setting, compliance incentives are not uniformly aligned across the population; instead, the best response depends on others' behavior and the current state of the epidemic. This qualitative shift results in higher infection levels in the forecasts compared to the 90-day training period.  

For the 150-day training window, Texas returns to Q1, but now in the Q1-Diminishing Oscillations regime. While the game again becomes cooperation-dominant, the oscillations of the infection level are diminishing, in contrast to the Q1-Oscillating TOC regime where the oscillations are amplified. Forecasts generated based on both 120- and 150-day training windows accurately capture the magnitude of infection levels.

\subsubsection{California}
\begin{figure*}[t]
\centering

% -------- Row 1 --------
\begin{subfigure}[b]{0.48\textwidth}
    \centering
    \includegraphics[width=\textwidth]{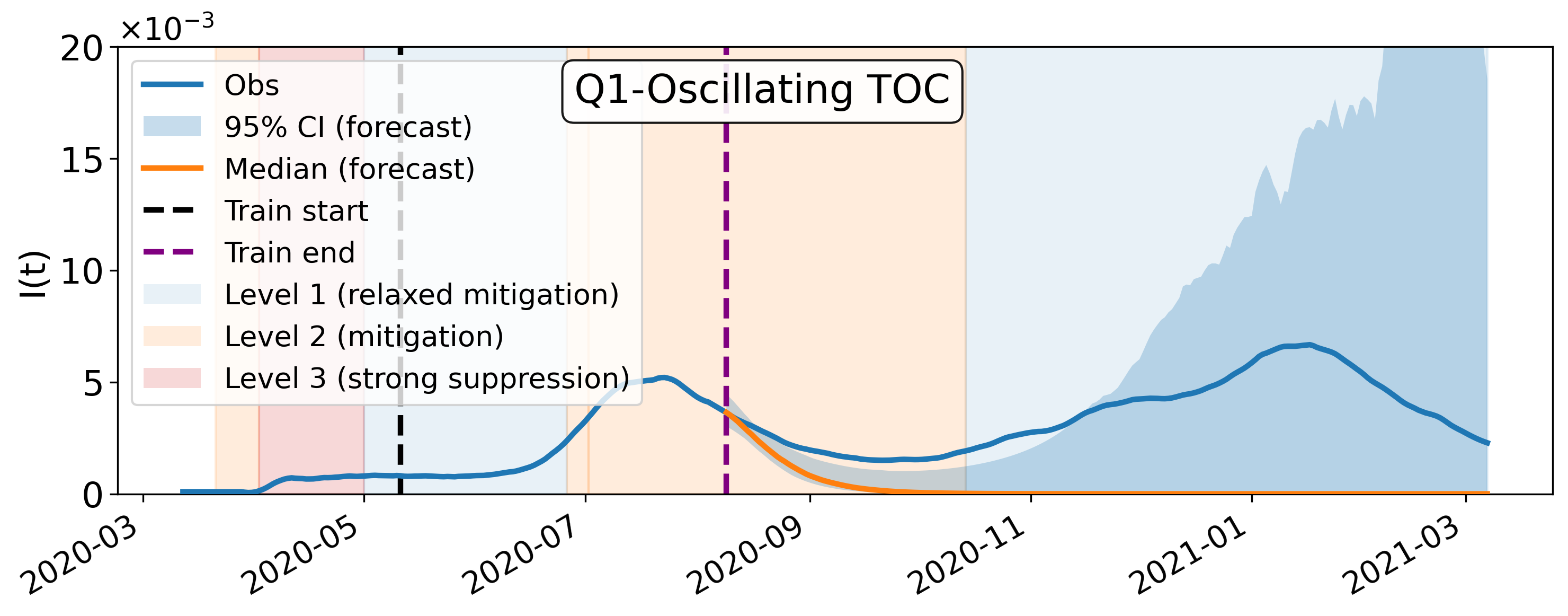}
    \caption{TX, 90-day training window}
    \label{fig:q1high}
\end{subfigure}
\hfill
\begin{subfigure}[b]{0.48\textwidth}
    \centering
    \includegraphics[width=\textwidth]{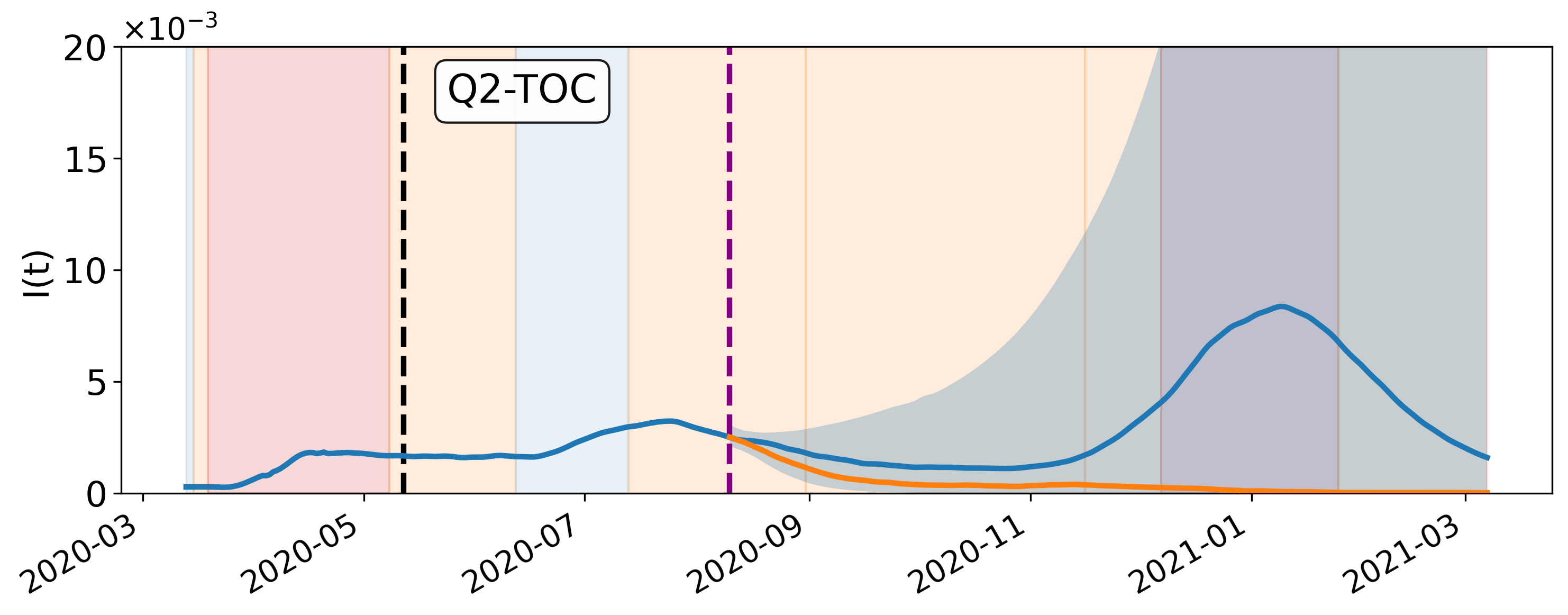}
    \caption{CA, 90-day training window}
    \label{fig:q2}
\end{subfigure}

\vspace{0.4cm}

% -------- Row 2 --------
\begin{subfigure}[b]{0.48\textwidth}
    \centering
    \includegraphics[width=\textwidth]{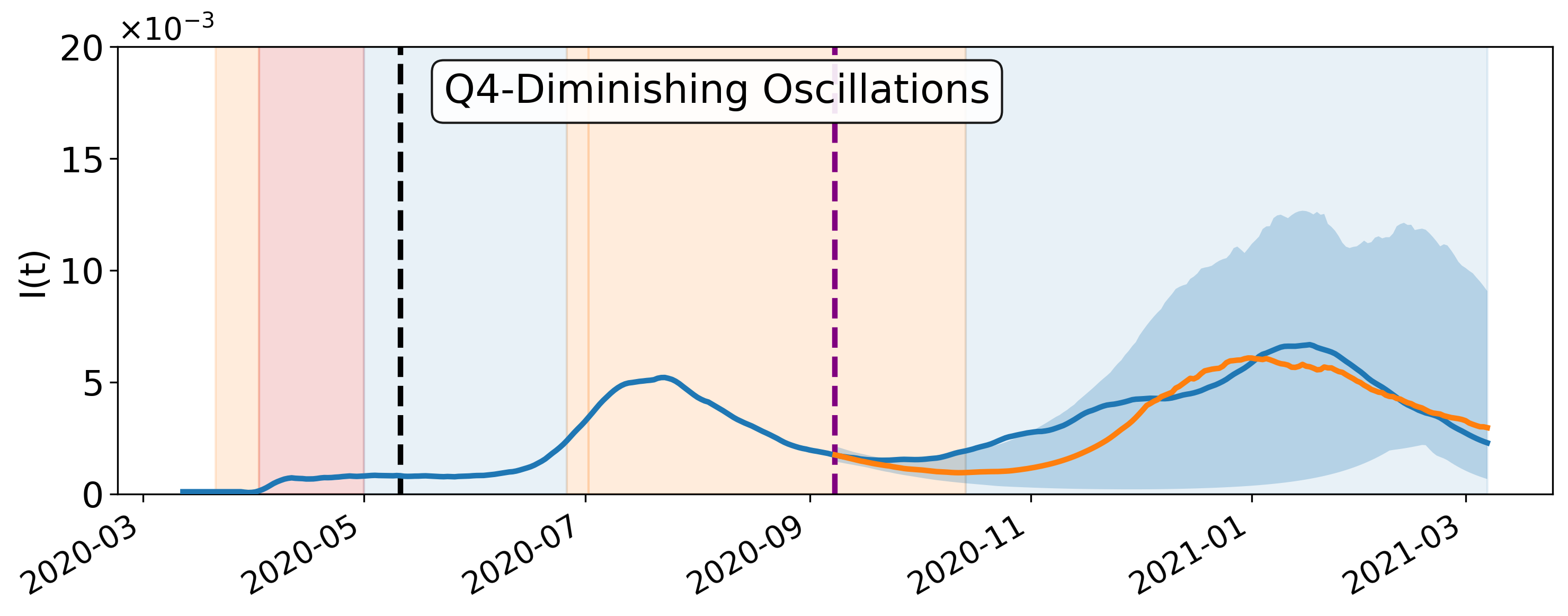}
    \caption{TX, 120-day training window}
    \label{fig:q3}
\end{subfigure}
\hfill
\begin{subfigure}[b]{0.48\textwidth}
    \centering
    \includegraphics[width=\textwidth]{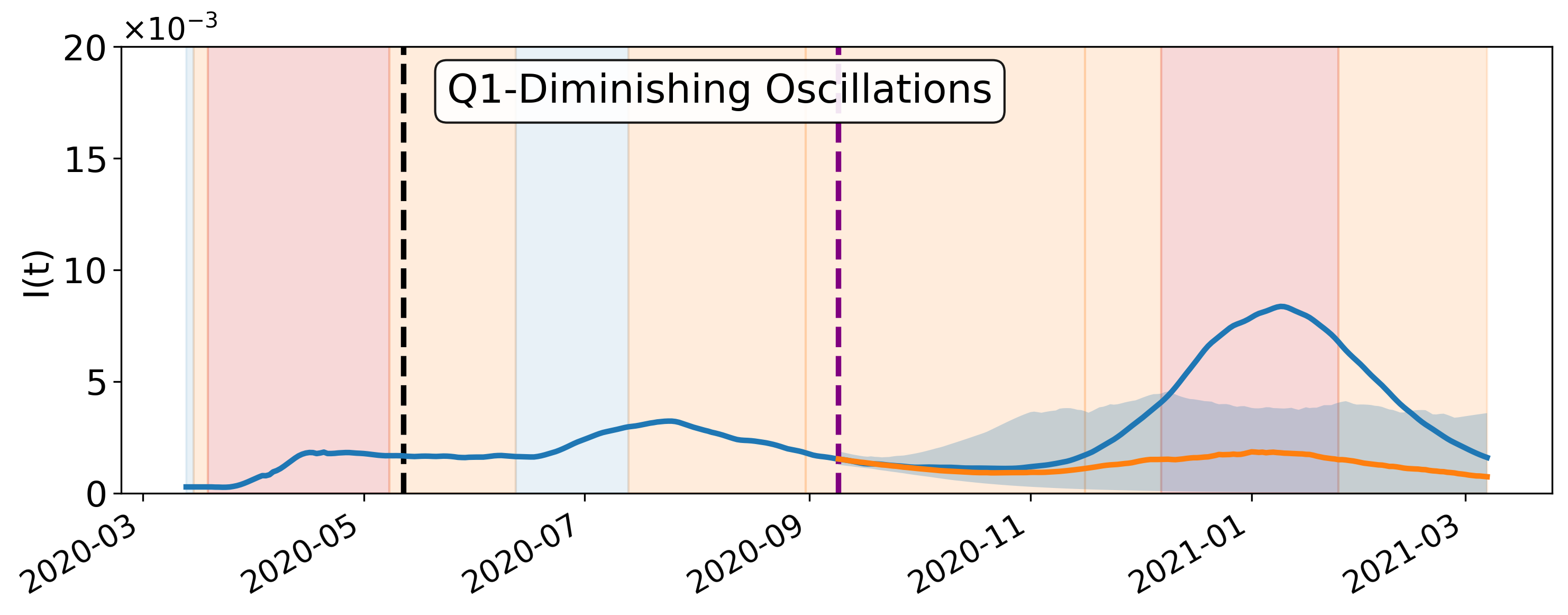}
    \caption{CA, 120-day training window}
    \label{fig:q4}
\end{subfigure}

\vspace{0.4cm}

% -------- Row 3 --------
\begin{subfigure}[b]{0.48\textwidth}
    \centering
    \includegraphics[width=\textwidth]{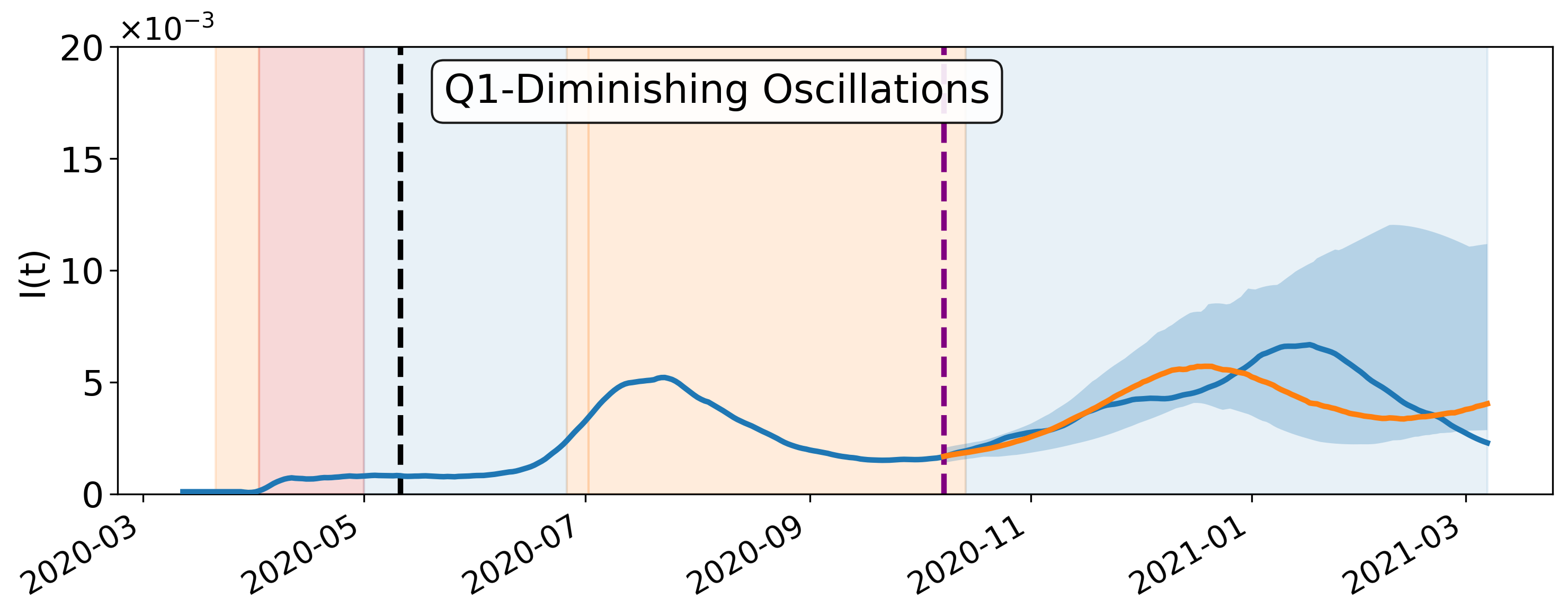}
    \caption{TX, 150-day training window}
    \label{fig:baseline}
\end{subfigure}
\hfill
\begin{subfigure}[b]{0.48\textwidth}
    \centering
    \includegraphics[width=\textwidth]{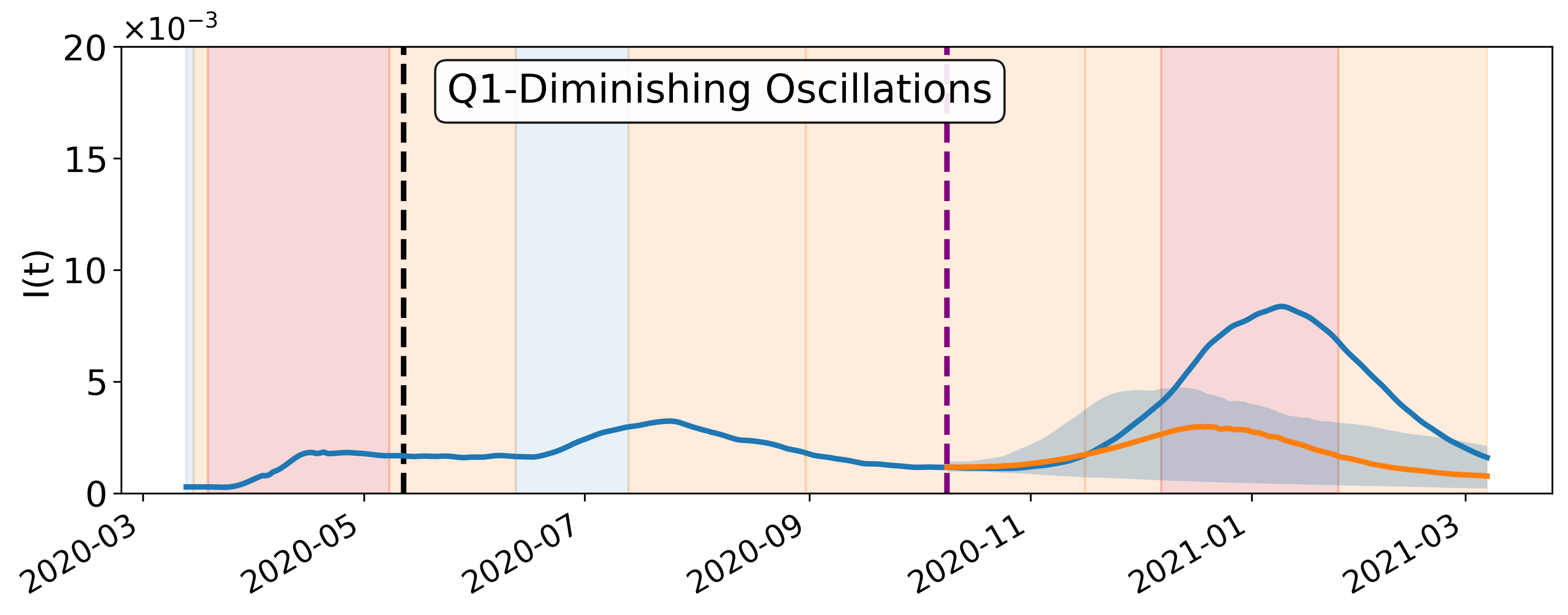}
    \caption{CA, 150-day training window}
    \label{fig:mc}
\end{subfigure}
\caption{Forecast results for Texas (TX) and California (CA) using 90-day (a--b), 120-day (c--d), and 150-day (e--f) training windows. We generate forecasts using the estimated payoff parameters $\hat{\bbdelta}$, initialized at the end of each training window. The training windows are marked with dashed vertical lines. The colored vertical regions show the intensity of mitigation policies in each state. Confidence intervals are constructed using Monte Carlo simulations, with shaded regions showing the 95\% confidence intervals. Forecasts for the following 2--3 months closely track the observed trajectories for both states.}
\label{fig:TXCA}
\end{figure*}

The model predictions remain within the confidence intervals for approximately the first 60--90 days, but they fail to capture the timing and magnitude of the second peak thereafter---see Fig.~\ref{fig:TXCA} (b,d,f). One possible explanation is that mitigation policies were implemented heterogeneously across regions within California, while the model learns a single aggregate behavioral response at the state level. As a result, the estimated payoff structure may not fully capture local variations in compliance and policy enforcement. This mismatch leads to confidence intervals that do not fully cover the observed peak, particularly during periods of rapid change. Despite this limitation, the model still captures the overall trend during the initial phase and provides reliable short-term forecasts.

Under the 90-day training window, California is classified in the Q2-TOC regime, where $\delta_{RT_0} > 0$ and $\delta_{SP_0} < 0$. This corresponds to a setting in which cooperation is not strictly dominant and mutual defection remains a Nash equilibrium in a high-infection environment. The inferred payoff matrices suggest that incentives to comply are present, but not strong enough to sustain coordinated behavior across the population.

With the 120-day and 150-day training windows, California transitions to Q1-Diminishing Oscillations, where both $\delta_{RT_0} > 0$ and $\delta_{SP_0} > 0$---see Fig.~\ref{fig:regime_tx_ca_az_co}. This indicates a cooperation-dominant game in $A_0$, in which compliance becomes more consistently beneficial across the population. Compared to the 90-day window, this training period places greater emphasis on extended periods of moderate mitigation (Level 2). The resulting regime implies that although oscillations may occur, they are damped over time rather than amplified. The forecasts after the 120-day and 150-day training windows are slightly better at capturing the general shape of the second wave, and the confidence intervals become more consistent with the observed trajectory. 

\begin{figure*}
    % -------- Row 4 --------
\begin{subfigure}[b]{0.48\textwidth}
    \centering
    \includegraphics[width=\textwidth]{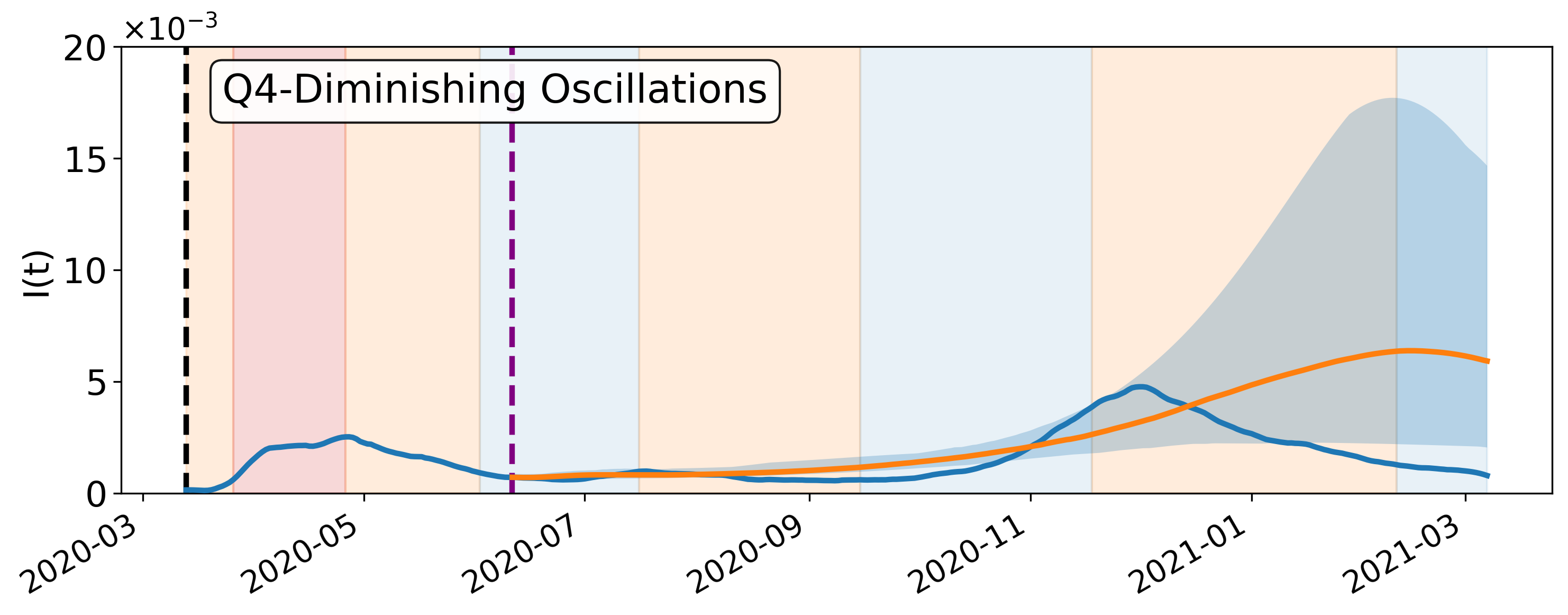}
    \caption{CO, 90-day training window}
    \label{fig:baseline}
\end{subfigure}
\hfill
\begin{subfigure}[b]{0.48\textwidth}
    \centering
    \includegraphics[width=\textwidth]{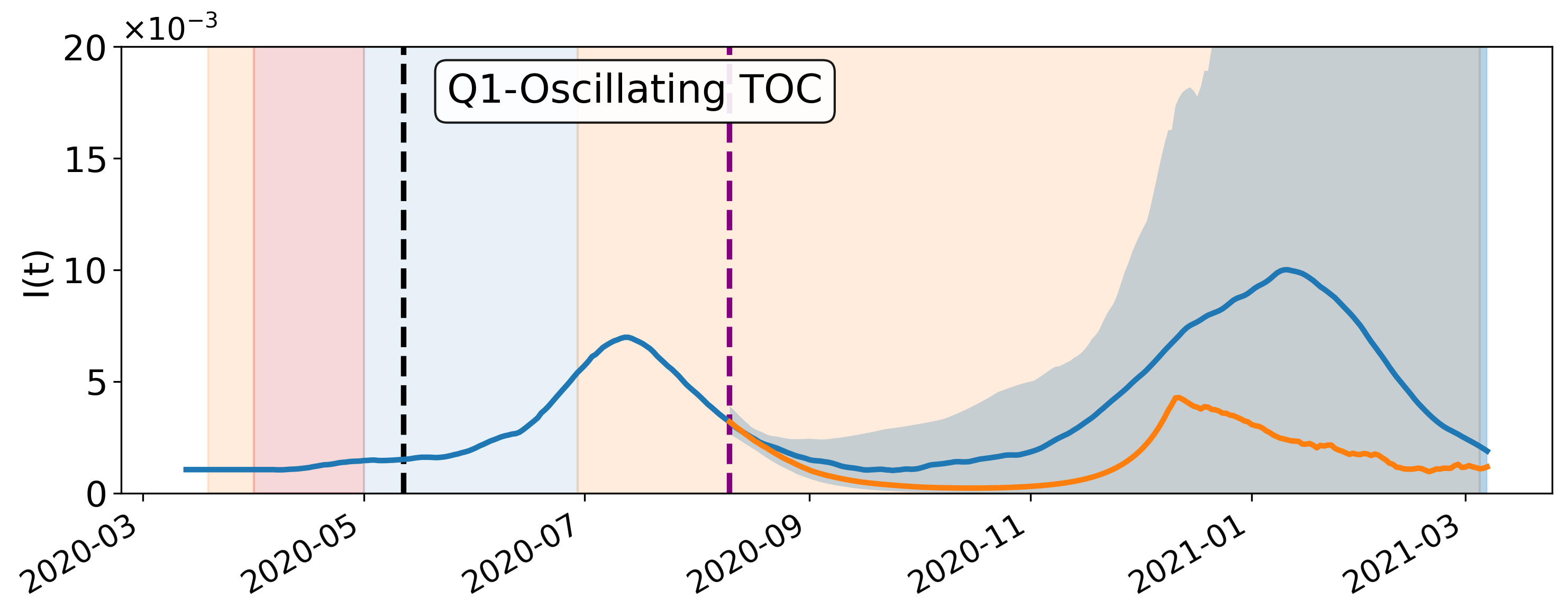}
    \caption{AZ, 90-day training window}
    \label{fig:mc}
\end{subfigure}

% -------- Row 5 --------
\begin{subfigure}[b]{0.48\textwidth}
    \centering
    \includegraphics[width=\textwidth]{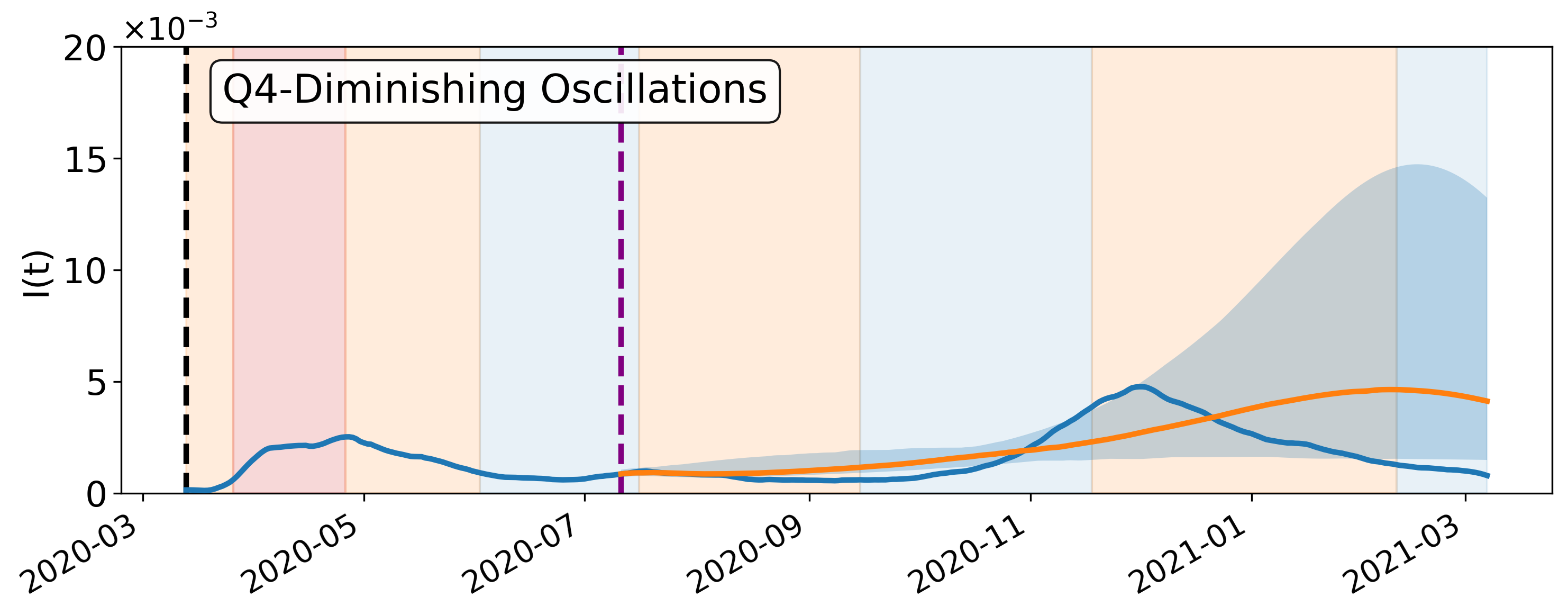}
    \caption{CO, 120-day training window}
    \label{fig:baseline}
\end{subfigure}
\hfill
\begin{subfigure}[b]{0.48\textwidth}
    \centering
    \includegraphics[width=\textwidth]{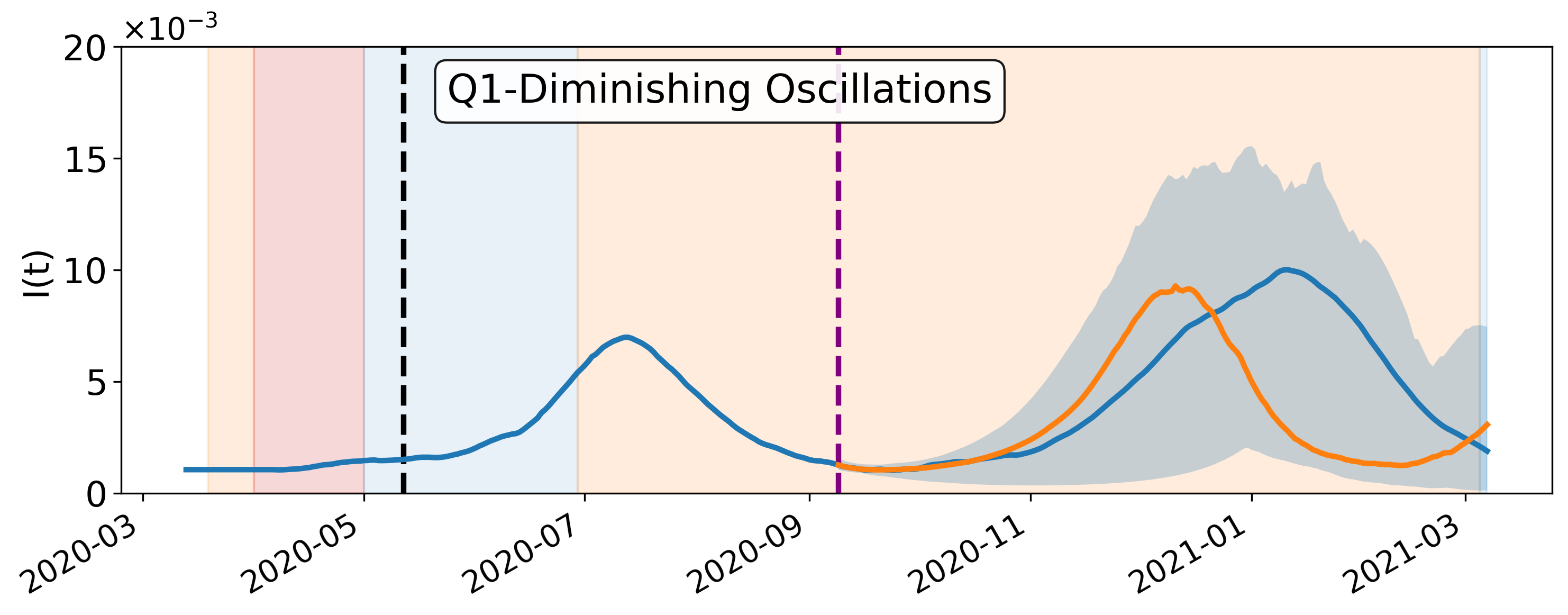}
    \caption{AZ, 120-day training window}
    \label{fig:mc}
\end{subfigure}

% -------- Row 6 --------
\begin{subfigure}[b]{0.48\textwidth}
    \centering
    \includegraphics[width=\textwidth]{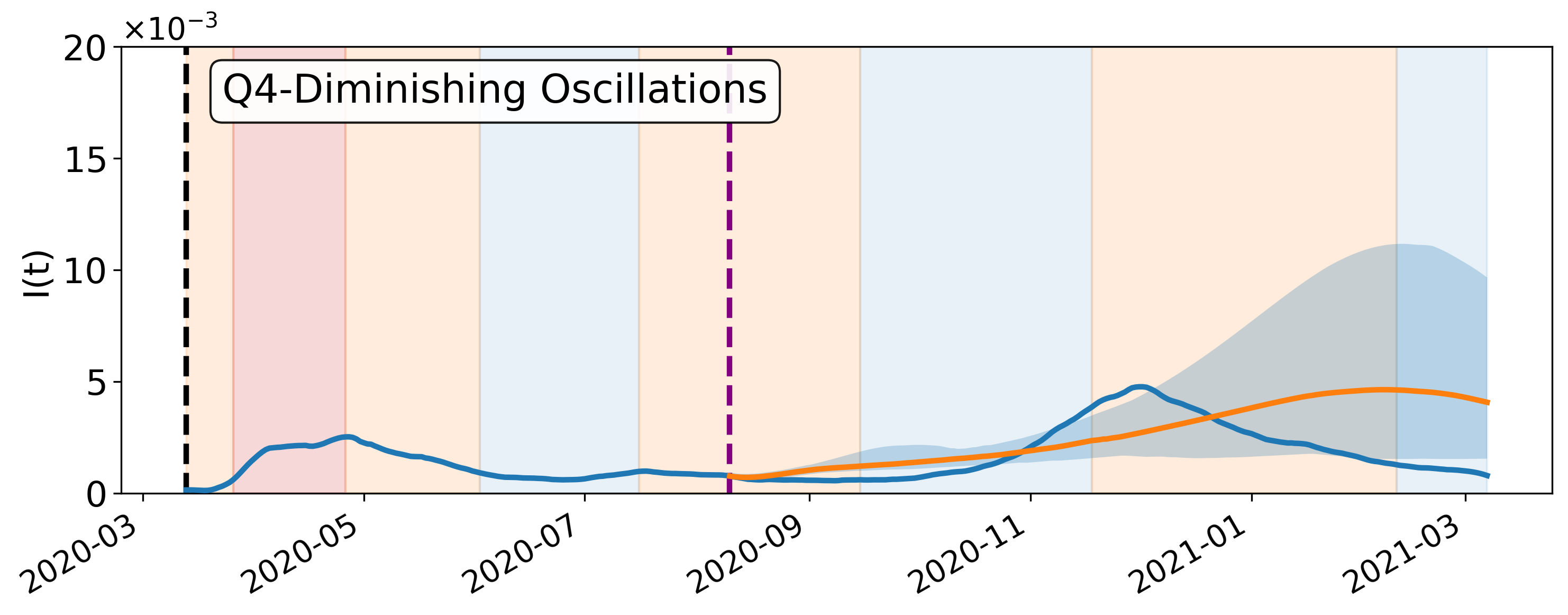}
    \caption{CO, 150-day training window}
    \label{fig:baseline}
\end{subfigure}
\hfill
\begin{subfigure}[b]{0.48\textwidth}
    \centering
    \includegraphics[width=\textwidth]{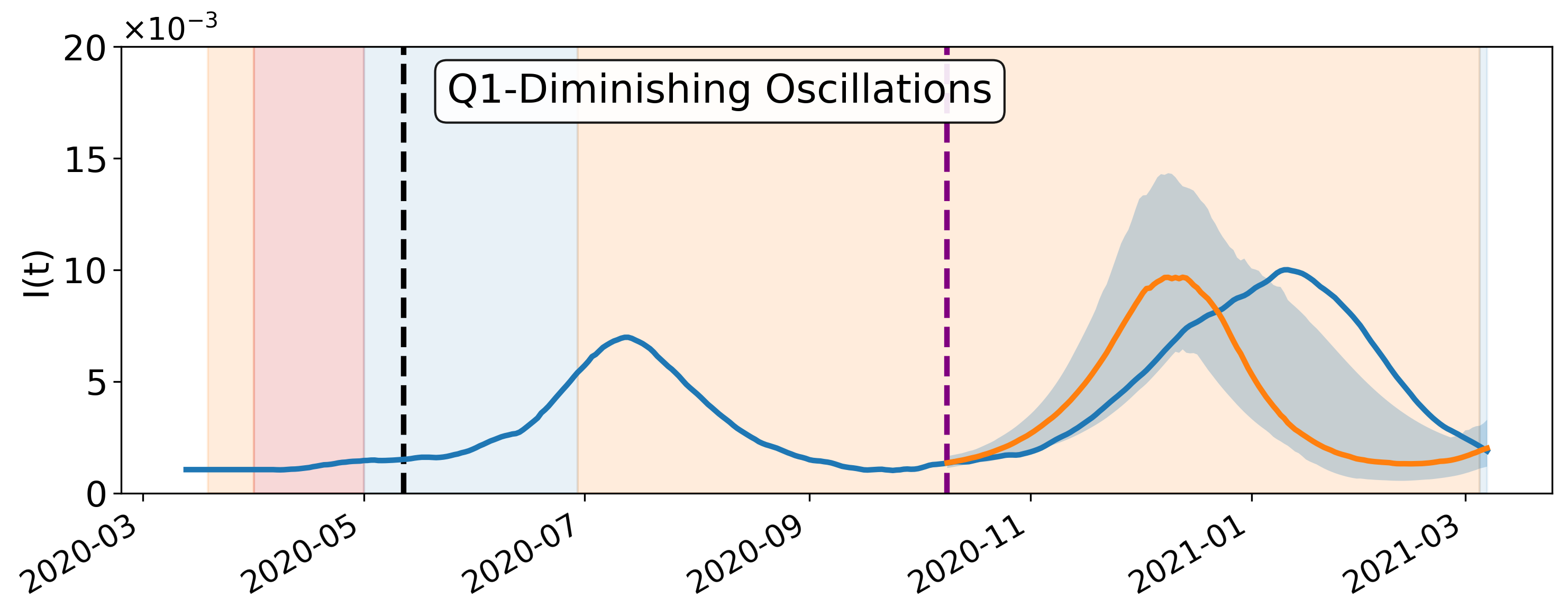}
    \caption{AZ, 150-day training window}
    \label{fig:mc}
\end{subfigure}
\caption{Forecast results for Colorado (CO) and Arizona (AZ) using 90-day (a--b), 120-day (c--d), and 150-day (e--f) training windows.}
\label{fig:COAZ}
\end{figure*}

\subsubsection{Colorado}
In Colorado, the first peak occurs relatively early; therefore, we start the training window at the beginning of the data collection period---see Fig.~\ref{fig:COAZ} (a,c,e). Following this initial peak, the infection trajectory remains relatively flat compared to other states until the second peak, and the overall dynamics exhibit low variability, with smaller peak magnitudes. The forecasts successfully capture this low-amplitude trend for approximately the first five months and maintain consistency with the observed data over the forecasting horizon. The confidence intervals consistently cover the observed trajectory, indicating that the model provides stable predictions even in scenarios with less pronounced epidemic peaks.

\begin{figure}[h]
    \centering
    \includegraphics[width=0.5\textwidth]{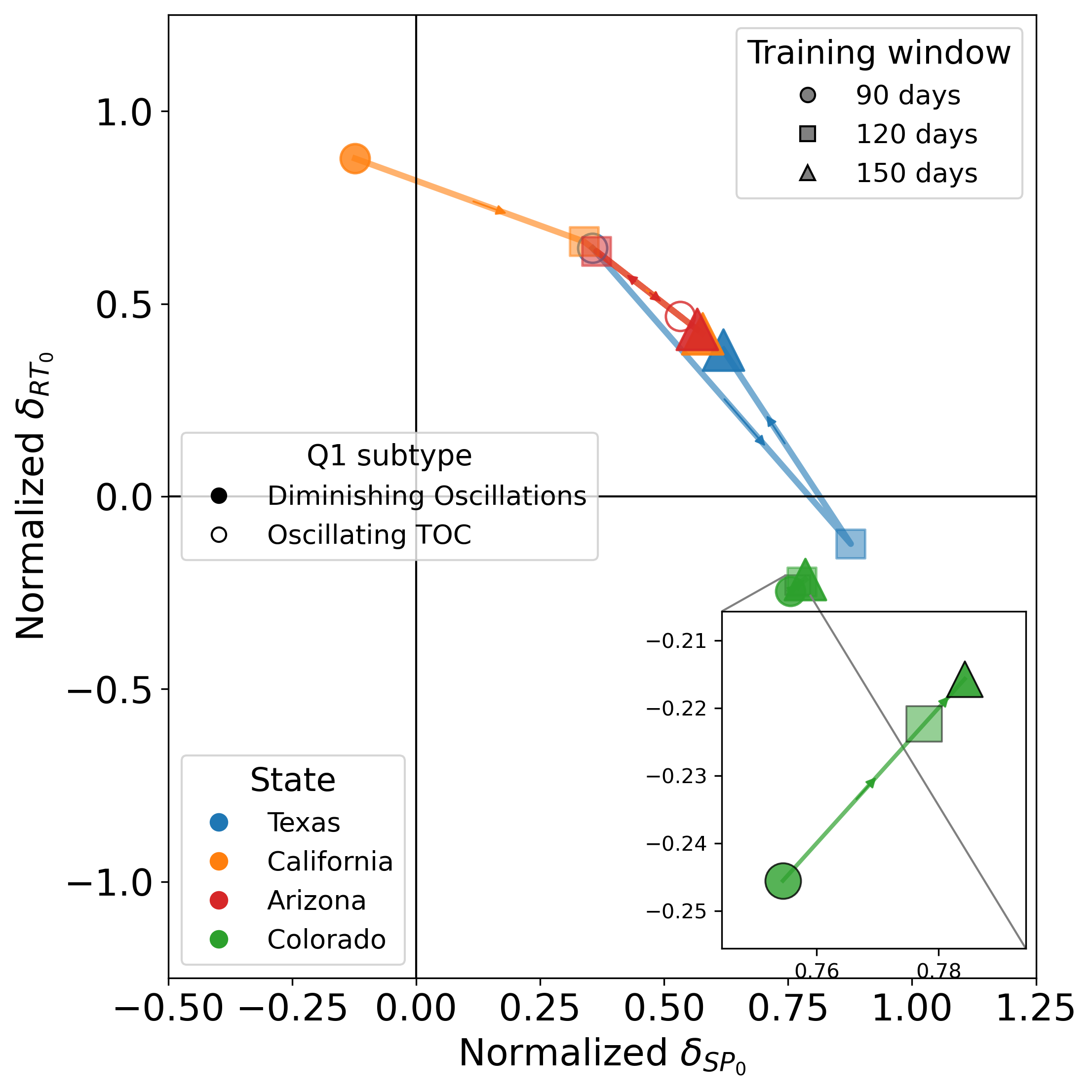}
    \caption{Each point represents the component-wise median payoff-gap estimate $(\delta_{SP_0}, \delta_{RT_0})$ obtained from 200 Monte Carlo simulations for a given state and training window. Lines connect the trajectory across training horizons for each state. In the Q1 regime, filled markers indicate diminishing oscillations, while unfilled markers indicate oscillating TOC. For visualization, the payoff-gap pairs are normalized by their $\ell_1$ norm, i.e., $(\delta_{SP_0}, \delta_{RT_0}) \mapsto (\delta_{SP_0}, \delta_{RT_0}) / (|\delta_{SP_0}| + |\delta_{RT_0}|)$, to place all estimates on a comparable scale. This is particularly necessary  as the magnitude of the estimates, especially for Colorado, is substantially larger. The corresponding raw values are reported in Table ~\ref{tab:delta_raw}.
}
    \label{fig:regime_tx_ca_az_co}
\end{figure}

Across all training windows (90, 120, and 150 days), Colorado is consistently classified as Q4-Diminishing Oscillations, characterized by $\delta_{RT_0} < 0$ and $\delta_{SP_0} > 0$. This corresponds to an anti-coordination game in the high-infection environment. This regime is associated with diminishing oscillatory dynamics, where infection levels may fluctuate but are not strongly amplified. Consistent with this interpretation, the model captures a relatively small peak and predicts a gradual, low-amplitude evolution thereafter, resulting in a relatively flat trajectory. While the magnitude of the payoff gaps decreases slightly as the training window increases, the qualitative regime classification remains unchanged---see Fig.~\ref{fig:regime_tx_ca_az_co}. In particular, even as longer training windows incorporate more prolonged exposure to relaxed mitigation levels (Level 1), the resulting behavioral regime remains consistent. In addition, the payoff-gap magnitudes are significantly larger than those of other states, implying a faster behavioral adaptation in the replicator dynamics--see Appendix~\ref{app:normalization} and Table~\ref{tab:delta_raw}. The magnitude value affects the rate of change of the cooperation level without altering the qualitative regime classification.

Overall, Colorado stands out as the most stable case among the four states, with all training horizons yielding the same Q4-Diminishing Oscillations regime. In game-theoretic terms, this indicates that the population consistently operates under an anti-coordination structure, and this inferred regime is consistent with the relatively flat epidemic trajectory.

\subsubsection{Arizona}
The model provides accurate forecasts for Arizona, capturing both the overall trend and peak behavior of the epidemic---see Fig.~\ref{fig:COAZ} (b,d,f). The confidence intervals remain well aligned with the observed data for up to approximately 60–90 days. Although the model predicts the peak slightly earlier than observed, it accurately captures the peak magnitude.

Across all training windows, Arizona remains in Q1, with $\delta_{RT_0}>0$ and $\delta_{SP_0}>0$, indicating that cooperation is the dominant action in $A_0$. However, the qualitative dynamics within Q1 change with the training horizon. The 90-day estimate is classified as Q1-Oscillating TOC, whereas the 120- and 150-day estimates are classified as Q1-Diminishing Oscillations. Thus, although the cooperation-dominant game structure is preserved, incorporating longer training periods changes the relative payoff structure sufficiently to shift the predicted dynamics from oscillatory resurgence to diminishing oscillations. These longer windows also incorporate greater exposure to Level 2 mitigation, providing additional information about behavioral responses under sustained mitigation.

\subsection{Comparison with the Standard SEIR Model}
\label{sec:seir_comparison}

\begin{figure*}[t]
\centering

\begin{subfigure}[b]{0.48\textwidth}
    \centering
    \includegraphics[width=\linewidth]
    {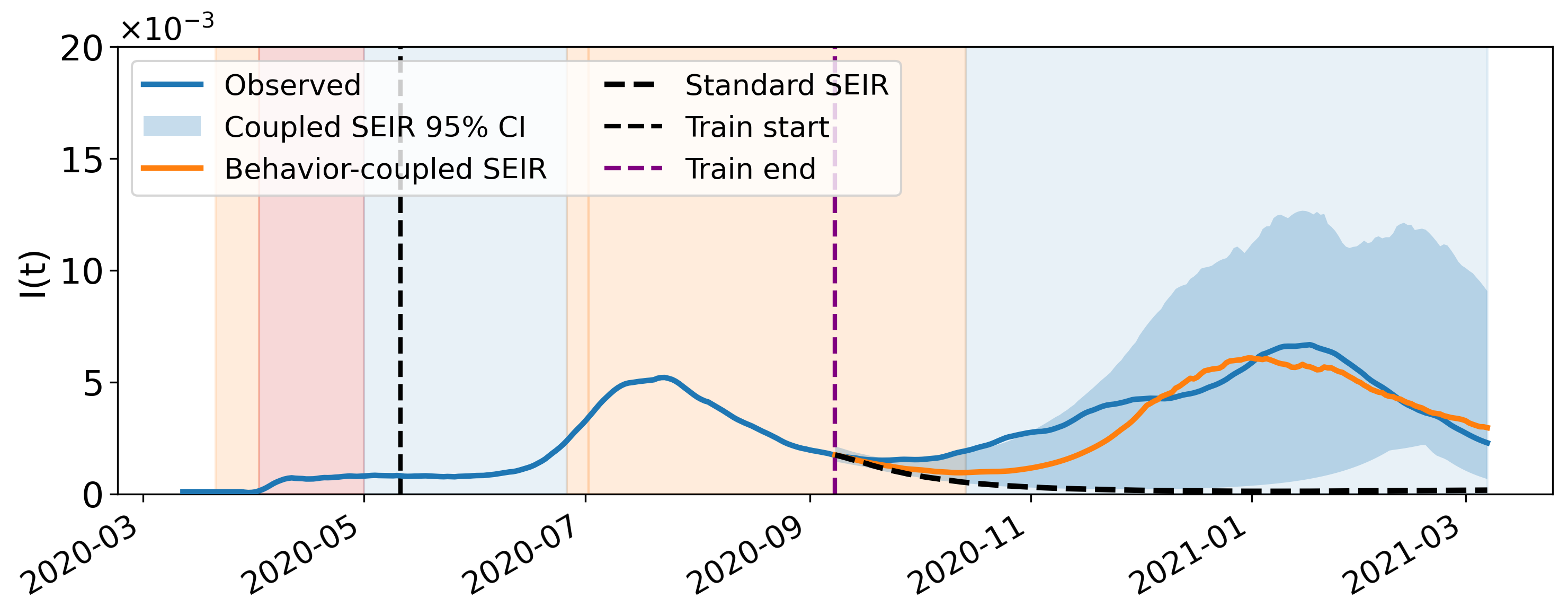}
    \caption{Texas}
    \label{fig:seir_tx}
\end{subfigure}
\hfill
\begin{subfigure}[b]{0.48\textwidth}
    \centering
    \includegraphics[width=\linewidth]
    {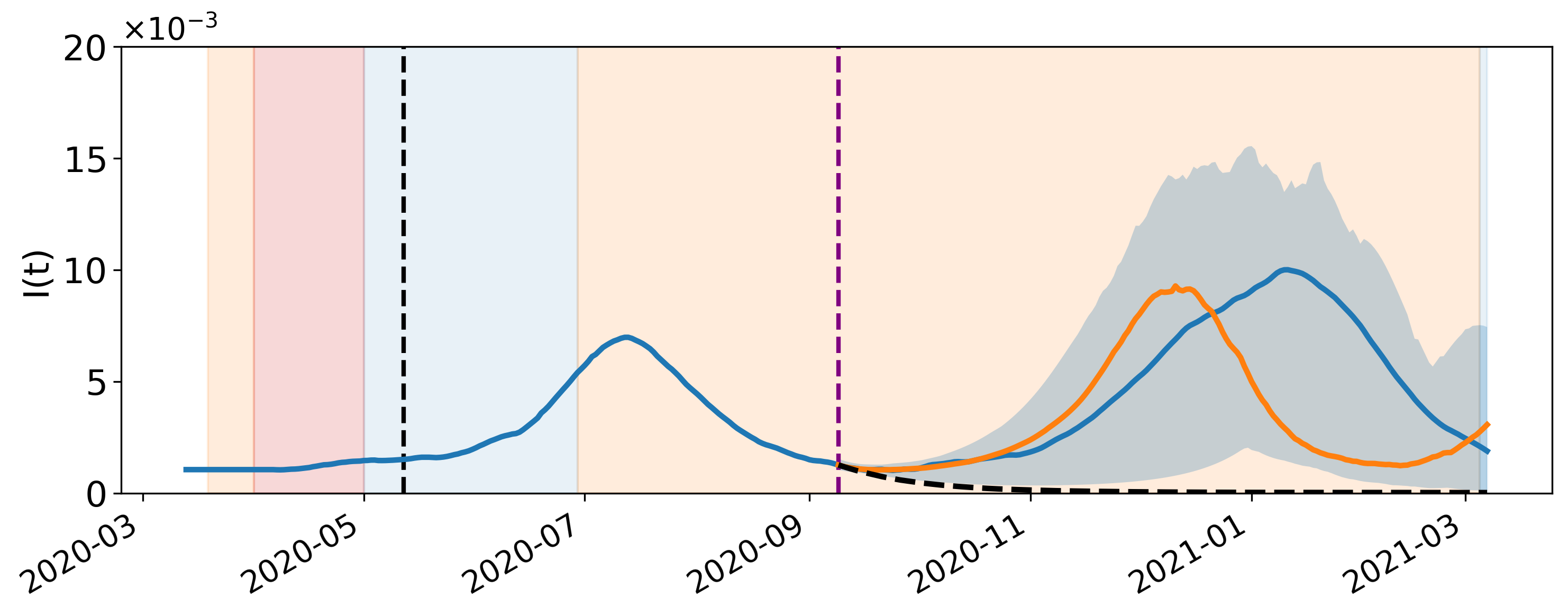}
    \caption{Arizona}
    \label{fig:seir_az}
\end{subfigure}

\vspace{0.3cm}

\begin{subfigure}[b]{0.48\textwidth}
    \centering
    \includegraphics[width=\linewidth]
    {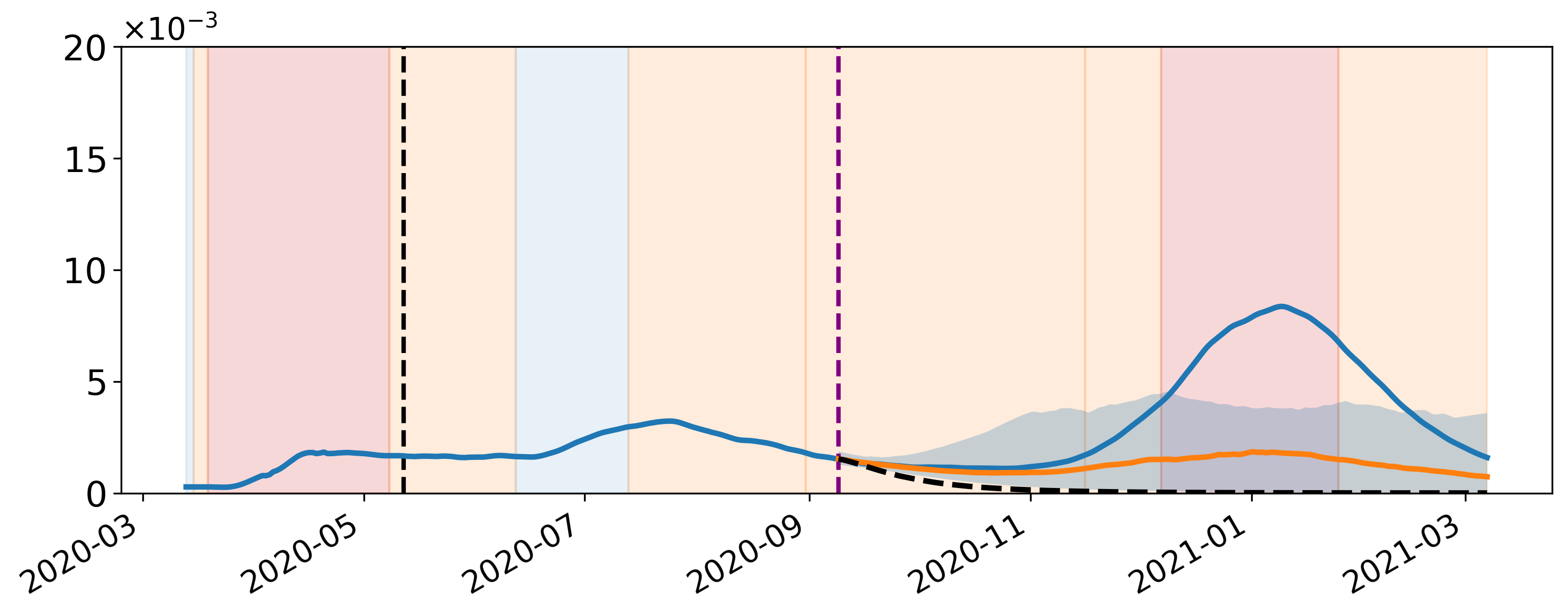}
    \caption{California}
    \label{fig:seir_ca}
\end{subfigure}
\hfill
\begin{subfigure}[b]{0.48\textwidth}
    \centering
    \includegraphics[width=\linewidth]
    {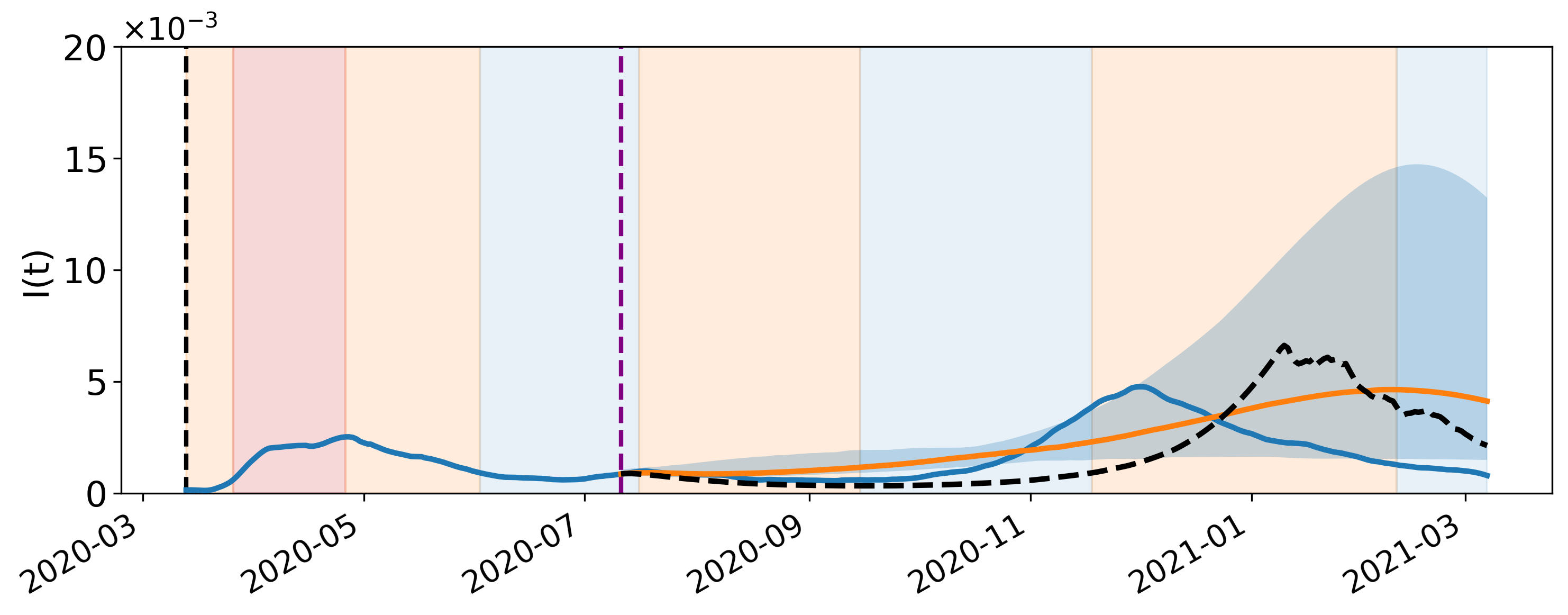}
    \caption{Colorado}
    \label{fig:seir_co}
\end{subfigure}

\caption{
Comparison of the behavior-coupled SEIR model and the standard seasonal
SEIR model for Texas, Arizona, California, and Colorado using the
120-day training window. The legend in panel (a) applies to all panels.
}
\label{fig:seir_comparison}
\end{figure*}

To quantify the forecasting contribution of endogenous behavioral feedback, we compare the proposed behavior-coupled SEIR model with an uncoupled seasonal SEIR model under the same epidemiological parameterization. Both models use the same reconstructed infection trajectory and, for each Monte Carlo realization, the same sampled disease-related parameters. The comparison is shown for the representative 120-day training window in Fig.~\ref{fig:seir_comparison}, while Table~\ref{tab:seir_rmse} reports the 60- and 90-day forecast errors.

\begin{table}[t]
\centering
\caption{
Out-of-sample forecast RMSE for the behavior-coupled and standard
seasonal SEIR models using the 120-day training window.
RMSE values are reported in units of $10^{-4}$.
Values are medians across Monte Carlo realizations, with interquartile
ranges in parentheses.
}
\label{tab:seir_rmse}

\small
\begin{tabular}{lccc}
\toprule
State
& Horizon 
& Behavior-coupled SEIR
& Standard SEIR \\
\midrule

TX
& 60 days
& $9.83\;(6.19,\,13.05)$ & $14.82\;(12.89,\,15.95)$ \\

TX
& 90 days
& $12.21\;(8.22,\,19.74)$ & $24.36\;(22.28,\,25.37)$ \\

AZ
& 60 days
& $5.12\;(2.38,\,8.14)$ & $11.62\;(10.83,\,12.14)$ \\

AZ
& 90 days
& $15.57\;(10.71,\,23.00)$
& $26.69\;(26.05,\,27.01)$ \\

CA
& 60 days
& $3.92\;(2.38,\,5.53)$
& $7.46\;(6.43,\,8.07)$ \\

CA
& 90 days
& $11.23\;(6.49,\,13.87)$
& $16.30\;(15.57,\,16.62)$ \\

CO
& 60 days
& $2.69\;(2.25,\,4.28)$ & $2.84\;(1.79,\,3.89)$ \\

CO
& 90 days
& $4.28\;(3.50,\,5.95)$ & $4.02\;(2.56,\,5.10)$ \\
\bottomrule
\end{tabular}
\end{table}

For Texas, Arizona, and California, the behavior-coupled model substantially improves forecast accuracy relative to the standard seasonal SEIR model. Based on the median RMSE values, the reduction in forecast error ranges from approximately $34\%$ to $56\%$ at the 60-day horizon and from approximately $31\%$ to $50\%$ at the 90-day horizon (Table~\ref{tab:seir_rmse}).

The improvement is most pronounced in Arizona, where the median 60-day RMSE decreases from $11.6\times10^{-4}$ for the standard SEIR to $5.12\times10^{-4}$ for the behavior-coupled model, corresponding to an approximately $56\%$ reduction. Texas shows a particularly large improvement at the 90-day horizon, with the median RMSE decreasing from $24.4\times10^{-4}$ to $12.2\times10^{-4}$. California also exhibits lower forecast errors at both horizons, although the coupled model underestimates the magnitude of the later winter peak.

Colorado differs from the other states. At the 60-day horizon, the two models have comparable performance, with median RMSEs of $2.69\times10^{-4}$ and $2.84\times10^{-4}$ for the behavior-coupled and standard SEIR models, respectively. At the 90-day horizon, the standard SEIR achieves a slightly lower median RMSE, $4.02\times10^{-4}$ compared with $4.28\times10^{-4}$ for the behavior-coupled model.

\section{Discussion and Conclusion} \label{sec:conclusion}

We considered a coupled SEIR-replicator dynamics model for forecasting epidemic spread. The model integrates disease dynamics with endogenous  behavioral responses. Beyond forecasting, the framework enables the estimation of population-level incentives underlying compliance decisions. A key contribution of this work is the ability to infer the payoff structure that governs behavior.  

The proposed forecasting approach captures epidemic trends in both synthetic data and infection trajectories inferred from hospitalization data. When applied to data from Arizona, California, Colorado, and Texas, the model is able to capture short- to medium-term dynamics, including features such as peak timing and magnitude in several cases. 
A further advantage of the proposed approach is that it leverages disease parameters from the literature and accounts for parameter uncertainty through perturbations around nominal values. Despite this uncertainty, the resulting forecasts remain stable and fall within reasonable confidence ranges. These results suggest that the framework is robust to model and parameter uncertainty, while providing interpretable behavioral and epidemiological insights into how individuals respond to infection risk and adjust their actions. Moreover, it can be readily adapted to different epidemic settings and applied to other infectious diseases and environments.

The comparison with the standard seasonal SEIR model clarifies the forecasting contribution of endogenous behavioral feedback. In Texas, Arizona, and California, the standard SEIR model predicts a rapid decline in infections after the forecast origin and fails to reproduce the subsequent resurgence over the relevant forecast period. In contrast, the behavior-coupled model allows changes in the behavioral state $x(t)$ to modulate transmission through $\beta(t)=\beta_0(t)(1-x(t))$, allowing the effective transmission rate to adjust endogenously as population behavior changes. This additional feedback mechanism substantially improves forecast accuracy in these three states, with reductions in median RMSE ranging from approximately $31\%$ to $56\%$ across the 60- and 90-day horizons.

The benefit of behavioral coupling is not uniform across all settings. Colorado provides an important counterexample. The two models show comparable performance at 60 days, while the standard SEIR achieves a slightly lower median RMSE at the 90-day horizon. Unlike the other states, seasonal transmission value alone generates a subsequent epidemic wave in Colorado, reducing the incremental short-horizon forecasting benefit of behavioral feedback. Beyond the evaluated 60- and 90-day horizons, both model trajectories also exhibit errors in the timing of the winter peak. Taken together, these results suggest that endogenous behavioral feedback is particularly valuable when seasonal transmission alone is insufficient to reproduce changes in epidemic growth, whereas its incremental forecasting benefit may be smaller when seasonal forcing already captures the dominant variation in transmission.

The model has limitations in capturing heterogeneity in population behavior. In particular, different age groups and sub-populations may exhibit distinct risk perceptions and behavioral responses. These effects could be incorporated by introducing group-specific behavioral states and payoff structures. While such extensions would increase the number of parameters to be estimated, they may provide detailed and actionable insights into population-level incentives.

Another limitation is the treatment of external drivers of behavior. The framework aims to capture the evolution of population behavior based on payoff structures given disease prevalence, and does not explicitly account for other factors, such as socioeconomic conditions, government mandates, lockdown orders, or enforcement efforts of mitigation policies (e.g., mask-wearing). As a result, the relationship between mandates and infection is indirect, and mediated through behavioral responses, which introduces additional uncertainty, particularly in long-term forecasts. At the same time, these inferred behavioral responses provide a quantitative proxy for changes in population behavior, which may reflect the aggregate effect of such external factors. Incorporating policy-aware or exogenous drivers of behavior to better capture these effects is an important direction for future work.

\appendices
% !TEX root = main.tex

\section{Prior results: Qualitative dynamics of behavior-coupled SEIR model} \label{Append: A}
The behavior-coupled SEIR system exhibits a wide range of possible dynamical outcomes, and their properties have been studied in previous work \cite{khazaei2021disease}. A summary of their qualitative dynamics is given in this section. The SEIR model considered here has no re-infection risk, as the loss of immunity to COVID is estimated to be at least 6 months, and we consider data sets to be approximately this time length. Thus, the only fixed point of the dynamics is the infection-free equilibrium $(i^* = 0)$.

\begin{figure}
    \centering
    \includegraphics[scale=0.3]{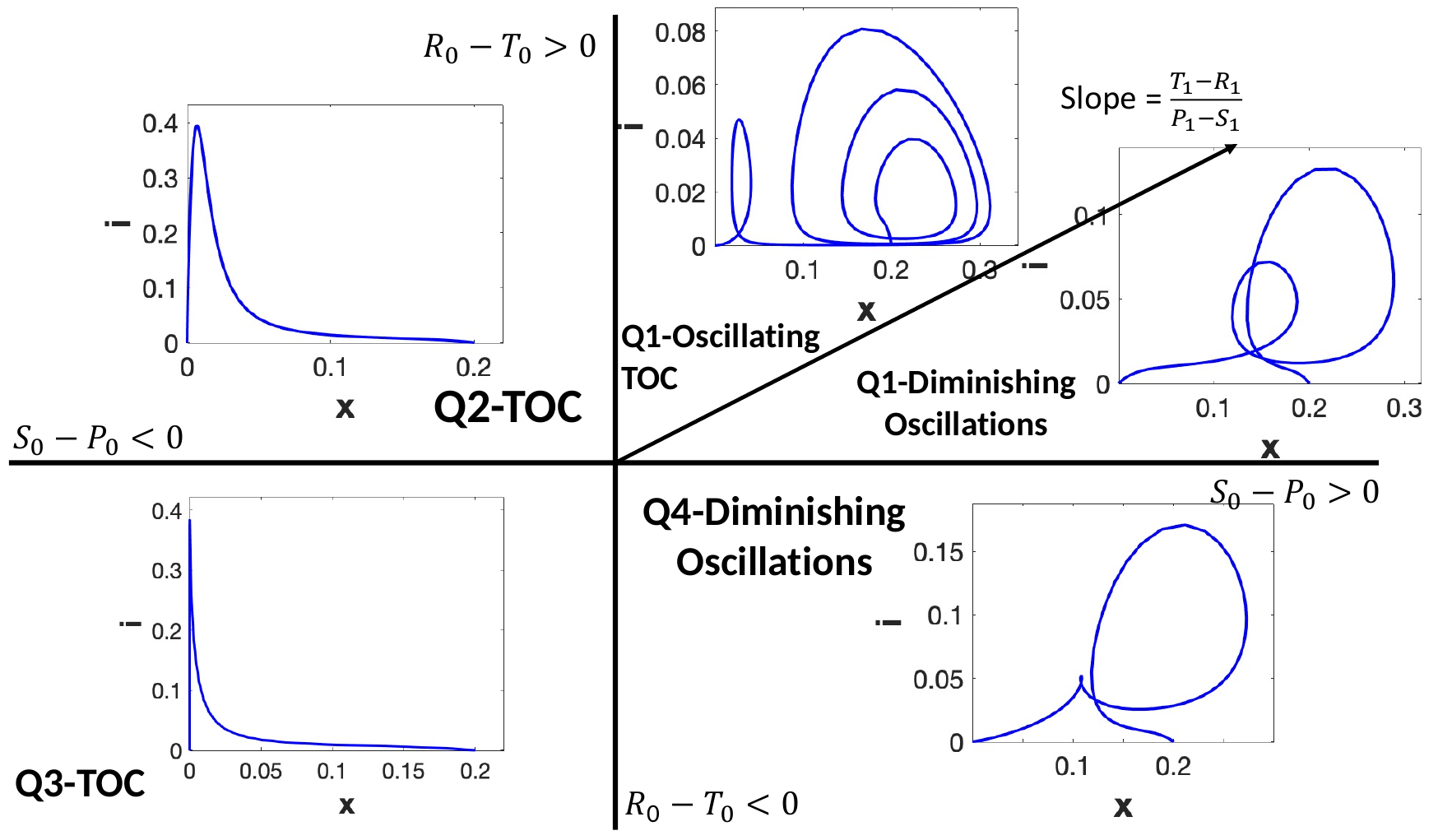}
    \caption{\small The range of possible dynamics that the behavior-coupled SEIR model can exhibit, parameterized by the payoff parameters in the $A_0$ matrix. The inset figures in each quadrant illustrate sample trajectories with respect to the two states $(x(t),i(t))$. Here, we set $A_1  =  \begin{bmatrix} 0.5 & 1 \\ 1 & 1.25  \end{bmatrix},\epsilon = 2,  s(0)=0.99,e(0)=0.01,i(0)=0,r(0)=0,x(0)=0.7,\frac{\beta_0}{\alpha}=5, \eta = 1$. The exact parameters on $A_0$ in each sample trajectory may be found in \cite{khazaei2021disease}.} 
    \label{fig:AllDynamics}
\end{figure}

Since all infections eventually die out in this model, all agents also eventually disregard social distancing guidelines ($x^* = 0$). However, the infection prevalence over time can display different transient dynamics. Depending on the payoff structure of the $A_0$ matrix, the SEIR dynamics can thoroughly be classified. Whenever mutual defection is a Nash equilibrium in $A_0$ ($\delta_{SP_0} :=S_0 - P_0 \leq 0$) the dynamics yield a single concentrated peak of infections, followed by exponential decay to zero. These are labeled as the top and bottom-left ``TOC" regimes in Figure \ref{fig:AllDynamics}.  When mutual defection is not a Nash equilibrium of $A_0$ ($\delta_{SP_0} > 0$), then the dynamics yield infectious oscillations, whose peaks are not as severe as the peak in the TOC regimes. When $A_0$ is an anti-coordination game ($\delta_{RT_0} :=R_0 - T_0 \leq 0$), the oscillations of $i_t$ are diminishing in amplitude (bottom-right Figure \ref{fig:AllDynamics}). When cooperation is the dominant action of $A_0$ ($\delta_{RT_0} > 0$), the oscillations are either still diminishing or initially growing in amplitude. The condition for oscillations growing in amplitude is
\begin{equation}\label{eq:eq_ratio}
    \frac{\delta_{RT_1}}{\delta_{SP_1}} < \frac{\delta_{RT_0}}{\delta_{SP_0}}
\end{equation}
and this regime is labeled as ``Oscillating TOC" in the top-right of Figure \ref{fig:AllDynamics}.

The coupled model with re-infection risk has also been studied \cite{khazaei2021disease}. In this scenario, there is a loss of immunity rate $\mu>0$ indicating flow from $r$ to $s$ compartments, and the model is termed SEIRS. Here, the $r$ state dynamics reads $\dot{r} = \alpha i - \mu r$. Under this model, it was shown that  either the infection-free equilibrium ($i^*=0$) is stable, or an endemic equilibrium is stable where a constant fraction of infected individuals persists. Similar dynamics were also identified in behavior-coupled SIS models, where conditions for the existence of stable limit cycles were derived \cite{frieswijk2022mean,paarporn2023sis}. Based on the payoff parameters, we classify the behavior regimes as follows.
\begin{itemize}
    \item Q2-TOC: $\delta_{RT_0}>0$, $\delta_{SP_0}<0$
    \item Q3-TOC: $\delta_{RT_0}<0$, $\delta_{SP_0}<0$
    \item Q4-Diminishing Oscillations: $\delta_{RT_0}<0$, $\delta_{SP_0}>0$
    \item Q1-Diminishing Oscillations: $\delta_{RT_0}>0$, $\delta_{SP_0}>0$, $\frac{\delta_{RT_0}}{\delta_{SP_0}} <\frac{\delta_{RT_1}}{\delta_{SP_1}}$
    \item Q1-Oscillating TOC: $\delta_{RT_0}>0$, $\delta_{SP_0}>0$, $\frac{\delta_{RT_0}}{\delta_{SP_0}} >\frac{\delta_{RT_1}}{\delta_{SP_1}}$
    \item Q1-Boundary: $\delta_{RT_0}>0$, $\delta_{SP_0}>0$, $\frac{\delta_{RT_0}}{\delta_{SP_0}} =\frac{\delta_{RT_1}}{\delta_{SP_1}}$
\end{itemize}

\section{TX, CA, CO, and AZ Data: Mandates and Disease Parameters} \label{app_real}

\subsection{Mandate Intensity Classification}\label{sec_mandates}
Using publicly available policy archives~\cite{husch_covid}, we classify each state's COVID-19 policies into three levels: \textbf{Level~3 (Strong Suppression)}---statewide or regional stay-at-home (SAH) orders, gathering bans, and closure of non-essential businesses; \textbf{Level~2 (Mitigation)}---strong restrictions short of SAH orders (indoor business shutdowns, mask mandates, reopening rollbacks, or high-restriction tiers); and \textbf{Level~1 (Relaxed Mitigation)}---phased reopening with capacity limits and partial lifting of restrictions. Table~\ref{tab:mandates} summarizes the level assignment over time for each state.

\begin{table*}[t]
\centering
\caption{Representative policy actions used to assign mandate levels by state.
SAH: stay-at-home. California entries are a statewide simplification and differ
in Los Angeles County.}
\label{tab:mandates}
\renewcommand{\arraystretch}{1.2}
\begin{tabular}{@{}l p{4.6cm} p{4.6cm} p{4.6cm}@{}}
\hline
\textbf{State} & \textbf{Level 3 (Strong)} & \textbf{Level 2 (Mitigation)} & \textbf{Level 1 (Relaxed)}\\
\hline
Texas      & Apr.\ 2020 statewide SAH order & Rollbacks (GA-28), mask mandate (GA-29), surge capacity limits & Phased reopening (GA-32/34), capacity limits and mask mandate lifted\\
Arizona    & Mar.--Apr.\ 2020 statewide SAH order & Reopening pauses, summer-surge business closures & Reopening with lifted occupancy limits\\
Colorado   & Mar.--Apr.\ 2020 statewide SAH order & Statewide mask mandate, COVID Dial winter mitigation, rollbacks & Safer-at-Home, Dial 2.0/3.0 reopening\\
California & Mar.\ 2020 SAH (EO N-33-20), Dec.\ 2020 Regional SAH & Indoor shutdowns, Purple Tier (Blueprint for a Safer Economy) & Tiered reopening (Red/Orange/Yellow)\\
\hline
\end{tabular}
\end{table*}

\subsection{Disease Parameter Selection} \label{parasel}
The nominal values of the disease-related parameters follow published COVID-19 estimates, and the $\pm 20\%$ ranges in Table~IV are sensitivity ranges that we impose rather than literature-derived confidence intervals. In each Monte Carlo realization, $\alpha$, $\eta$, $\kappa$, $\beta_0^s$, and $\beta_0^w$ are drawn from Gaussian distributions with standard deviation equal to $10\%$ of the nominal values, and draws outside $\pm 20\%$ of the nominal values are clipped to the nearest bound; the integer hospitalization delay $\tau$ is obtained by rounding a Gaussian draw with mean $9$ days and standard deviation $0.9$ days and restricting it to $\{7,8,9,10\}$. Winter peaks of SARS-CoV-2 transmission in temperate regions motivate the seasonal form of $\beta_0(t)$~\cite{gavenvciak2022seasonal,wiemken2023seasonal}; its magnitude is guided by reported reproduction numbers~\cite{flaxman2020estimating}, and the state-specific values in Table~IV are modeling choices that reflect climatic differences (see the table note). The hospitalization delay $\tau=9$ days lies within the reported median intervals from symptom onset to hospital admission, which range from $3$ to $10.4$ days depending on age and are longest for ages 20--60~\cite{faes2020time}; we thereby assume that entry into the infectious compartment approximately coincides with symptom onset. The hospitalization fraction $\kappa=0.07$ is based on a population-based estimate of the infection hospitalization rate  of $6.86\%$ in the non-congregate population of Connecticut~\cite{mahajan2021sars}. The recovery rate $\alpha=0.15$ ($1/\alpha\approx 6.7$ days) is consistent with evidence that infectiousness declines within about a week of symptom onset~\cite{he2020temporal}. Finally, $\eta=0.143$ ($1/\eta\approx 7$ days) is set from a pooled incubation period of about $6.6$ days~\cite{wu2022incubation}; using the incubation period as the latent period of the $E\to I$ transition neglects pre-symptomatic transmission.

\subsection{Additional Results on Payoff Gap Magnitudes and Normalization} \label{app:normalization}
We introduce a normalization factor defined as $c := |\delta_{SP_0}| + |\delta_{RT_0}|.$ Using this, the replicator dynamics in \eqref{eq_rep_expanded} can be equivalently rewritten as
\begin{align}
\dot x = c \, x(1-x)& \cdot \Bigg[
x \theta \frac{\delta_{RT_1}}{c} + x (1-\theta) \frac{\delta_{RT_0}}{c} \nonumber\\
&+ (1-x) \theta \frac{\delta_{SP_1}}{c} + (1-x) (1-\theta) \frac{\delta_{SP_0}}{c}
\Bigg].
\end{align}
The normalized payoff gaps determine the qualitative regime classification, while the normalizing factor $c$ affects only the time scale of the behavioral adaptation. Consequently, normalization preserves the qualitative trajectories of the system.

\begin{table}[h]
\centering

\caption{Median estimates of the payoff-gap parameters
$(\delta_{RT_0}, \delta_{SP_0})$ before normalization across states and training windows.}

\label{tab:delta_raw}

\begin{tabular}{c c c c c}
\hline
State & Training window & $\delta_{RT_0}$ & $\delta_{SP_0}$ & $c$ \\
\hline

\multirow{3}{*}{Texas}
& 90  & 1.302  & 0.720  & 2.022 \\
& 120 & -0.214 & 1.521  & 1.735 \\
& 150 & 0.534  & 0.869  & 1.403 \\
\hline

\multirow{3}{*}{California}
& 90  & 2.306  & -0.324 & 2.630 \\
& 120 & 0.735  & 0.376  & 1.111 \\
& 150 & 0.397  & 0.542  & 0.940 \\
\hline

\multirow{3}{*}{Arizona}
& 90  & 0.664  & 0.758  & 1.422 \\
& 120 & 0.776  & 0.444  & 1.220 \\
& 150 & 0.505  & 0.659  & 1.164 \\
\hline

\multirow{3}{*}{Colorado}
& 90  & -73.158  & 224.789 & 297.947 \\
& 120 & -27.397 & 95.813  & 123.210 \\
& 150 & -21.707 & 78.920  & 100.627 \\
\hline

\end{tabular}

\par\footnotesize
\textit{Note:} Values are rounded to three decimal places.

\end{table}

\bibliographystyle{IEEEtran}
\bibliography{TexFiles/Ref}

\end{document}